\documentclass[journal,onecolumn]{IEEEtran}
\usepackage[top=1in,left=1in,right=1in,bottom=1in]{geometry}
\usepackage{amsmath, mathrsfs,amsfonts}
\usepackage{amssymb}
\usepackage{color}
\usepackage{acronym}
\usepackage{graphicx}
\usepackage{epsfig}
\usepackage{subfigure}
\usepackage{graphicx}
\begin{document}
\title{Performance Analysis of Dipole Antennas Embedded in Core-Shell Spheres: A Green's Function Analysis}
\author{Shabnam~Ghadarghadr,\IEEEmembership{Student Member, IEEE,}, Mohsen F. Farahani, and Hossein Mosallaei,\IEEEmembership{Senior Member, IEEE}\\
Applied Electromagnetics Laboratory, \\
ECE Department, Northeastern University, Boston, MA, 02115}
\markboth{IEEE Transactions on Antennas and Propagation}%
\date{}
\maketitle
\begin{abstract}
The main goal of this work is to theoretically investigate the behavior of an electrically small antenna enclosed in a concentric sphere.
The Green's function analysis is applied to characterize the input impedance of a concentric resonator excited by a dipole located at
its center. The method of moments (MoM) with Galrekin's procedure is used to determine the current distribution over the source excitation and hence the input impedance. The behavior of quality factor (Q) and bandwidths of the antenna is studied with the use of input impedance as a function of frequency. We illustrate that by embedding a dipole antenna inside a core-shell structure, with magnetic shell and dielectric core, a Q as low as the Chu limit can be approached. The obtained observations demonstrate how a resonator composed of magnetic shells can provide electrically small antennas with high bandwidths performance.
\end{abstract}
\section{Introduction}
\quad Antennas with smaller physical size and wider bandwidths are desirable for mobile systems.
In recent years, considerable efforts have been devoted towards antenna miniaturization. The main challenge is to reduce the resonant frequency while maintaining other radiation features. According to Chu limit the ability of the antenna to radiate effectively depends on the total size of the antenna ($k_0a$ with $k_0$ being the wave number in the free space and $a$ being the radius of the smallest sphere enclosing the antenna) \cite{chu}.
To achieve an antenna with desired impedance characteristics, one can either design new topologies \cite{ref1,ref2} or use different substrates. \cite{ref4,hossein}. One of the most conventional methods for attaining desired miniaturization is to utilize high dielectric substrate, so that the size can be scaled down by $\sqrt{\epsilon_r}$. However, due to the strong coupling between the high dielectric material and the ground plane, a large amount of energy is trapped inside the high permittivity substrate, and a hence good radiator cannot be achieved. To reduce this coupling it has been suggested to use magneto-dielectric materials with moderate values of $\mu_r$ and $\epsilon_r$ \cite{ref4,hossein}. Note that the effect of the permeability is similar to adding some inductance to the input impedance of the antenna. Thus, the antenna provides more radiation to the outside region leading to a wideband small antenna \cite{ref4,hossein}. Nevertheless it should be mentioned that achieving a low-loss weak-frequency-dispersive magnetic material at the desired frequency is a major challenge. 

\quad Another approach for antenna miniaturization is to embed the antenna inside a negative permittivity (ENG) resonator. For instance, Stuart et al in \cite{stuart} excited an ENG sphere with a dipole feed to produce the appropriate polarization required for the resonance performance. They demonstrated that one can achieve a very small antenna by operating the sphere around the frequency where $\epsilon_r=-2$ region. Spherical geometry of the structure offers wide-band performance approaching the Chu limit. They showed that in this way Chu limit can be approached but not exceeded \cite{stuart,conf6,conf7,conf8,j3}. This will be very beneficial for the design of small-size antennas (the radius of sphere can be much smaller than the wavelength). Ziolkowski et al. have also demonstrated other novel small antennas designs utilizing combinations of ENG and MNG or Double Negative (DNG) metamaterials \cite{zil}. Spherical geometry of the structure ensures the low-quality factor impedance performance.

\quad As mentioned earlier, with the use of ENG , the quality factor around 1.5 times Chu limit can be achieved \cite{stuart}. In this paper we propose a novel approach to achieve smaller Q (around $1.08Q_{Chu}$) with the use of magnetic materials. The idea is to combine both electric and magnetic effects by using a core-shell to reduce the absorbed energy inside the resonator. It is worth mentioning that if the absorbed energy inside the resonator is negligible compared to the stored energy outside the resonator, the operating Q would be around the Chu limit. To do that end, we embed the antenna inside a core-shell resonator where the thin shell is made of magnetic materials and the core is filled with conventional dielectrics.
We shows that in this scenario a quality factor around $1.08Q_{Chu}$ can be achieved.

\quad In this paper, a rigorous treatment of a dipole fed a concentric spherical antenna is presented.
In such a structure, the energy first radiates from the dipole to the concentric resonator and then from the resonator to the free space.
To explore the behavior of small antennas embedded in multi-layer spheres, we utilize Green's function analysis. The Green's functions for the evaluation of the input impedance are derived and the method of moment (MoM) with Galerkin's procedure is used to determine the probe current from which the input impedance of the resonator is calculated. The desired Green's function is presented as sum of the Green's function associated with the dipole and the Green's function associated with the scattering of the multi-layered sphere. For the dipole, the traditional cylindrical Green's function is used and then expanded in term of spherical waves for matching the boundary conditions. By having the incident Green's function (the dipole Green's function) the Green's function associated with the discontinuity of the concentric sphere can be achieved. The scattered coefficients are obtained by imposing the boundary conditions.
To verify the accuracy of our technique, comparison between HFSS and MoM is applied leading to a very good agreement.

\quad We then, explore the behavior of small antennas embedded inside a spherical core-shell through input impedance analysis.
The bandwidth performance of the antenna when embedded inside a spherical resonators is investigated. Different cases, including substrate with high dielectric, magneto-dielectric and magnetic are considered. For each case, the quality factor is obtained and compared with the Chu limit.
We illustrate that by designing an optimized core-shell with magnetic shell and dielectric core  quality factor as low as Chu limit can be attained. From practical point of view, it is worth highlighting that although it is hard to obtain a  high $\mu$ material in GHz, it is relatively easy to achieve them when the thickness is thin.

\quad The structure of this paper is as follows. In section II, the Green's functions for deriving resonance frequencies and input impedance of the antenna are obtained. In section III numerical results which verifies our theory for miniaturizing the antenna are presented and finally we conclude everything in the section IV.

\section{Green's Function Analysis}

\begin{figure}[htbp]
  \centering
      \subfigure{\scalebox{0.4}{\epsfig{file=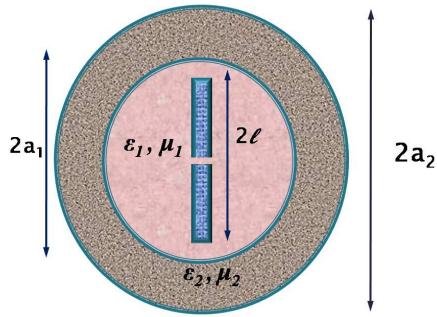}}}
  \caption{The antenna configuration, a dipole antenna inside a spherical core-shell resonator}
    \label{fig:dipolecore}
\end{figure}

\quad In this section, we represent a rigorous analysis of electromagnetic waves' radiation and scattering in the presence of a core-shell sphere. The required dyadic Green's functions are derived and expressed in the form of an infinite series of spherical eigen-modes. This series are convergent and hence can be truncated using a finite number of terms. Having the Green's function, we can simply find the electric field behavior and hence the input impedance.
The geometry of a dipole antenna inside a spherical core-shell is shown in Fig. \ref{fig:dipolecore}.
The inner and outer radii of the core-shell are $a_1$ and $a_2$ respectively. The material of the core has the permittivity of $\epsilon_1$ and permeability of $\mu_1$, where the shell is described with $\epsilon_2$ and $\mu_2$. The dipole of length $2l$ is located at the center of the core-shell. The source is considered to be a delta gap at the center of the dipole.

The goal is to obtain the input impedance, hence the z-directed electric field due to the z-direct current over the dipole should be calculated.
The z-directed electric field due to the z-directed current element is determined by first expressing the z-directed current as r-directed current.
The r-directed current can only generate $TM^{r}$ modes, since only electric vector potential needs to be evaluated. As mentioned earlier, the desired Green's function is the sum of the Green's function associated with the dipole inside a homogenous medium of $\epsilon_{r1},\mu_{r1}$ and the Green's function due to the discontinuity of the core-shell. In this work we refer to the Green's function associated with the discontinuity of the sphere as the scattering Green's function. In order to achieve the scattering Green's function, one has to first obtain the  incident Green's function (due to the dipole inside a homogenous medium). In this work, we first start with the electric vector potential Green's functions. Once the potential Green's functions are derived, the various field components \cite{harrington} and hence the input impedance as well as the radiation patterns can be evaluated.
To begin with we derived the solution for the r-directed Green's function due to the incident dipole.
The incident Green's function represent an infinitesimal source radiating in the unbounded medium of $\epsilon_{r1},\mu_{r1}$.
The solution for an r-directed electric source can be obtained by solving the following differential equation \cite{harrington}:
\begin{equation}
\label{eq:GrindEQ__2chap6}
\left(\nabla ^{2} +k^{2} \right)\frac{\mathbf{G}_{J_{r'} }^{A_{i} } }{r} =\frac{-1}{r} \frac{\delta \left(r-r^{'} \right)\delta \left(\theta -\theta ^{'} \right)\delta \left(\phi -\phi ^{'} \right)}{r^{2} \sin \theta }.
\end{equation}
Where $G_{J_{r'} }^{A_{i}}$ is the incident Green's function for the electric vector potential caused by an r-directed current. $\hat{J}_{n} \left(x\right),\hat{H}_{n}^{(2)} \left(x\right)$ are spherical Bessel and Hankel functions respectively and $P_{n}^{m} \left(x\right)$ is the associated Legendre function.
Therefore the r-directed Green's function $G_{J_{r'} }^{A_{i}}$ is:
\begin{subequations}
\begin{align}
&{{G}}^{A_i}_{J_{r'}}|_{r>r'}=\hat{r}\sum^n_{n=0}\frac{2n+1}{n(n+1)}P_n(\cos\zeta)\hat{J}_n(k_1r')\hat{H}^{(2)}_n(k_1r)\\
&{{G}}^{A_i}_{J_{r'}}|_{r<r'}=\hat{r}\sum^n_{n=0}\frac{2n+1}{n(n+1)}P_n(\cos\zeta)\hat{J}_n(k_1r')\hat{J}_n(k_1r)
\end{align}
\end{subequations}
It is worth noticing that ($r^{'} ,\theta ^{'} ,\phi ^{'} $) refers to source point while $\left(r,\theta ,\phi \right)$ refers to field point, and
\begin{equation}
\cos\zeta=cos\theta\cos\theta'+\sin\theta\sin\theta'\cos(\phi-\phi')
\end{equation}
For an antenna at the center $\theta'=0$ and hence $\cos\zeta=\cos\theta\cos\theta'$.
In the same way the scattered Green's function for the electric can be expresses in terms of spherical harmonics:
\begin{subequations}
\label{eq:greelayer0}
\begin{align}
&{\mathbf{G}}^{A_1}_{J_{r'}}=\hat{r}\sum^\infty_{n=0}C_{s1}(n,r',\theta')P_n(\cos\theta)\hat{J}_n(k_1r)\\
&{\mathbf{G}}^{A_2}_{J_{r'}}=\hat{r}\sum^\infty_{n=0}C_{s2}(n,r',\theta')P_n(\cos\theta)\left[\hat{J}_n(k_2r)+\hat{H}^{(2)}_n(k_3r)R_{23}(n)\right]\\
&{\mathbf{G}}^{A_3}_{J_{r'}}=\hat{r}\sum^\infty_{n=0}C_{s3}(n,r',\theta')P_n(\cos\theta)\hat{H}^{(2)}_n(k_3r)
\end{align}
\end{subequations}
where the indices $1,2,3$ refer to inside the core, inside the shell and outside the core-shell respectively. Note that, we assume a very thin dipole antenna, thus the structure is symmetric over the $\phi$ and  $\phi'$. This means that in the expression of incident and scattering Green's function there is no dependence on $\phi$ or $\phi'$. By matching the boundary conditions, which are the continuity of the electric and magnetic fields at $a_1$ and $a_2$, the unknown coefficients $C_{s1},C_{s2},C_{s3}$ and $R_{23}(n)$ can be found. Using the above, the total Green's functions in each layers for the electric vector potentials are defined as follows (for $r>r'$):
\begin{subequations}
\label{eq:greelayer}
\begin{align}
&{\mathbf{G}}^{A_1}_{J_{r'}}=\hat{r}\sum^n_{n=0}\frac{2n+1}{n(n+1)}P_n(\cos\zeta)\hat{J}_n(k_1r')\left[\hat{H}^{(2)}_n(k_1r)+R_{12}\hat{J}_n(k_1r)\right],\\
&{\mathbf{G}}^{A_2}_{J_{r'}}=\hat{r}\sum^n_{n=0}C_1(n)\frac{2n+1}{n(n+1)}P_n(\cos\zeta)\hat{J}_n(k_2r')\left[\hat{H}^{(2)}_n(k_2r)+R_{23}\hat{J}_n(k_2r)\right],\\
&{\mathbf{G}}^{A_3}_{J_{r'}}=\hat{r}\sum^n_{n=0}C_2(n)\frac{2n+1}{n(n+1)}P_n(\cos\zeta)\hat{J}_n(k_3r')\hat{H}^{(2)}_n(k_3r),
\end{align}
\end{subequations}
where $G^{A_i}_{J_r'}$ for $i\in\{1,2,3\}$ are the total Green's functions for the electric vector potential in layer $i$.
It is worth highlighting that $k_i$ for $i\in\{1,2,3\}$ refers to the wave number in each layer of the spherical core-shell. Also for a dipole located at the center of the core-shell,
\begin{equation}
\cos\zeta=\cos\theta\cos\theta'.
\end{equation}
As shown in \eqref{eq:greelayer} the Green's functions in each layer is related to the summation of $\hat{H}^{(2)}_n(kr)$ and $\hat{J}_n(kr)$. For the first layer $\hat{H}^{(2)}_n(k_1r)$ is associated with the dipole and $\hat{J}_n(k_1r)$ is associated with the effect of the core-shell. In the third layer, we have outgoing waves in an unbounded medium and hence $G^{A_3}_{J_r'}$ is proportional to $\hat{H}^{(2)}_n(k_3r)$.
The unknown coefficients of scattered fields in \eqref{eq:greelayer} are obtained by applying the boundary conditions which are the continuity of the tangential electric and magnetic fields at $r=a_1$ and $r=a_2$.
\begin{subequations}
\label{eq:gener}
\begin{align}
&R_{23}=\frac{\frac{k_{3} }{\epsilon _{3} } \hat{H}_{n}^{(2)} \left(k_2a_2\right)\hat{H}_{n}^{2'} \left(k_{3}a_2\right)-\frac{k_2}{\epsilon } \hat{H}_{n}^{(2)} \left(k_{3} a_2\right)\hat{H}_{n}^{(2)'} \left(k_2a_2\right)}{\frac{-k_{3} }{\epsilon _{3} } \hat{J}_{n} \left(k_2a_2\right)\hat{H}_{n}^{(2)'} \left(k_{3} a\right)+\frac{k_2}{\epsilon_2} \hat{H}_{n}^{(2)} \left(k_{3} a_2\right)\hat{J}'_{n} \left(k_2a_2\right)},\\
&R_{12}= \nonumber\\
&\frac{\frac{k_2}{\epsilon_2}\hat{H}^{(2)}_n(k_1a_1)\left[\hat{H}'^{(2)}_n(k2a_1)+R_{23}\hat{J}'_n(k2a_1)\right]
-\frac{k_1}{\epsilon_1}\hat{H}^{(2)}_n(k_1a_1)\left[\hat{H}'^{(2)}_n(k2a_1)+R_{23}\hat{J}'_n(k2a_1)\right]}
{\frac{k_2}{\epsilon_2}\hat{J}_n(k_1a_1)\left[\hat{H}'^{(2)}_n(k2a_1)+R_{23}\hat{J}'_n(k2a_1)\right]
-\frac{k_1}{\epsilon_1}\hat{J}_n(k_1a_1)\left[\hat{H}'^{(2)}_n(k2a_1)+R_{23}\hat{J}'_n(k2a_1)\right]}.
\end{align}
\end{subequations}
$R_{12}$ and $R_{23}$ are the generalized reflection coefficients in a multi-layer sphere \cite{chew}. It is worth highlighting that $R_{23}$ can be interpreted as the TM scattering coefficient of a simple sphere filled with $\epsilon_2$ and $\mu_2$ inside a homogeneous medium of $\epsilon_3$ and $\mu_3$. Once we obtained the generalized reflection coefficients other coefficients can be achieved:
\begin{subequations}
\label{eq:magcoef}
\begin{align}
&C_1(n)=\frac{\hat{H}^{(2)}_n(k_1a_1)+R_{12}\hat{J}_n(k_1a_1)}{\hat{H}^{(2)}_n(k_2a_1)+R_{23}\hat{J}_n(k_2a_1)}\\
&C_2(n)=C_1(n)\frac{\hat{H}^{(2)}_n(k_2a_2)+R_{23}\hat{J}_n(k_2a_2)}{\hat{H}^{(2)}_n(k_3a_2)}
\end{align}
\end{subequations}
By having all the unknown coefficients in \eqref{eq:greelayer}, and hence having the Green's function for the electric vector potential, the Green's function for all components of electric field can be attained \cite{harrington}. Let us define a dyadic Green's function for the electric field associated with the r-directed current.
\begin{equation}
\overline{\overline{G}}^E_{J_r}=G^{E_r}_{J_r}\hat{r}\hat{r}+G^{E_\theta}_{J_r}\hat{\theta}\hat{r}+G^{E_\phi}_{J_r}\hat{\phi}\hat{r}
\end{equation}
Where $G^{E_r}_{J_r}$ is the Green's function for the r-directed electric field derived from $G^A_{J_r}$. In the same manner $G^{E_\theta}_{J_r}$ and $G^{E_\phi}_{J_r}$ are Green's functions for the $\theta$-directed and $\phi$-directed components of the electric field respectively.
The electric field is then can be obtained solving the following integral equation:
\begin{align}
\label{eq:elecf}
&\mathbf{E} =\iint \nolimits _{S_{0} }\overline{\overline{G}}^{E}_{J_r}~.~\mathbf{J}_{r'}{\hat{r}} dS
\end{align}
Where $S_{0}$ is the surface on which the current is flowing and $J_{r}$ is the r-directed current.
The next step in obtaining the antenna input impedance which requires the z-directed electric field. We can extract the z-component of the electric field from \eqref{eq:elecf}. By comparing the the extracted field with the following equation the desired Green's function  for the evaluation of the
input impedance is thus obtained.
\begin{align}
\label{eq:elecf}
&\mathbf{E} =\iint \nolimits _{S_{0} }{{G}}^{E_z}_{J_z}~.~\mathbf{J}_{z'}{\hat{z}} dS
\end{align}
The z-directed Green's function for the electric field for a dipole located at the center of the core-shell is equal to:
\begin{align} \label{eq:GrindEQ__3chap6}
&{{G}}^{E_z}_{J_z}=\cos\theta\cos\theta'{{G}}^{E_r}_{J_r}
\end{align}
By having the z-directed Green's function for the electric field, the input impedance can be achieved. We use the input impedance as a function of frequency to extract antenna features such as bandwidths and quality factor.

\quad It is found that the incident solution (dipole field) is a slowly convergent series of Hankel functions and hence an excessive number of terms are required \cite{lin}. To avoid this difficulty, we use the well known Green's function of a cylindrical dipole with the reduced kernel \cite{harrington,lin}.
Combining the incident field and the scattering field, the z-directed Green's function due to the z-directed current can be written as
(note that for a thin dipole the $\phi ,\phi '$ variation can be ignored)
\begin{subequations}
\begin{align}
&\mathbf{G}_{J_{z } }^{E }=\mathbf{G}_{J_{z } }^{E_{i} } + \mathbf{G}_{J_{z} }^{E_{s} }\\
&\mathbf{G}_{J_{z } }^{E_{i} } =\hat{z}\frac{1}{j\omega \epsilon_1 } \left(\frac{\partial ^{2} }{\partial z^{2} } +k_1^{2} \right)\frac{e^{-jk\sqrt{\left(z-z'\right)^{2} +r_{1}^{2} } } }{4\pi \sqrt{\left(z-z'\right)^{2} +r_{1}^{2}}}\\
 &\mathbf{G}_{J_{z'} }^{E_{s } } = \hat{z}\frac{1}{4\pi k \omega \epsilon}
\frac{-\cos \theta \cos \theta '}{k r^{2} r'^{2} } \sum _{n=1}^{\infty }n\left(n+1\right)R_{12} P_{n}(\cos \theta \cos \theta ^{'})\hat{J}_{n} \left(r'\right) \hat{J}_{n} \left(kr\right)
\end{align}
\end{subequations}
Where $\mathbf{G}_{J_{z^{'} } }^{E_{i} }$ and $\mathbf{G}_{J_{z^{'} } }^{E_{s} }$ stand for the incident and scattering Green's functions for the electric field associated with the z-directed current.

\quad Finding the electric field Green's functions, the radiator's input impedance can be then
determined by applying the method of moment with Galerkin's procedure, and piecewise sinusoidal
(PWS) basis functions. Delta gap source model is considered for the feed. The unknown coefficients of
dipole current $(I_n)$ are obtained from the following matrix equation \cite{lin,j3,conf6,conf7,conf8}.
\begin{equation} \label{eq:mommatrix}
\left[Z_{mn} \right]\, \, \left[I_{n} \right]=\left[Z_{mn}^{i} +Z_{mn}^{s} \right]\, \, \left[I_{n} \right]=\left[V_{m} \right]
\end{equation}
where
\begin{subequations}
\begin{align}
&Z_{_{mn} }^{i} =-\int _{z}\int _{z'}f_{m} \left(z\right)\, \, \left(G_{J_{z'} }^{E_{i} } \right)  \, f_{n} \left(z'\right)\, dz\, dz'\\
&Z_{_{mn} }^{s} =-\int _{z}\int _{z'}f_{m} \left(z\right)\, \left(G_{J_{z'} }^{E_{s_{inside} } } \right)  \, f_{n} \left(z'\right)\, dz\, dz'x
\end{align}
\end{subequations}
$f_n(z)$ is the PWS basis function and is defined as follows for $z_n\leq z\leq  z_{n+1}$ \cite{lin}:
\begin{align}
&f_n(z)=\frac{sin{k(d-|z-z_{n+1}|)}}{\sin{kd}}\\
&z_n=-l+(n-1)d\\
&d=\frac{2l}{N+1}
\end{align}
With N being the number of segmentations. It should be mentioned that when the reduced kernel is used together with PWS modes, the
accuracy as well as the convergence of Z matrix, for a moderately thick dipole can be improved by using the two term
equivalent radius as discussed in \cite{lin}.
A computer program is developed to obtain the current distribution and the input impedance. Once the input impedance is obtained, the quality factor can be calculated as follows.

\begin{figure}[htbp]
  \centering
      \subfigure{\scalebox{0.33}{\epsfig{file=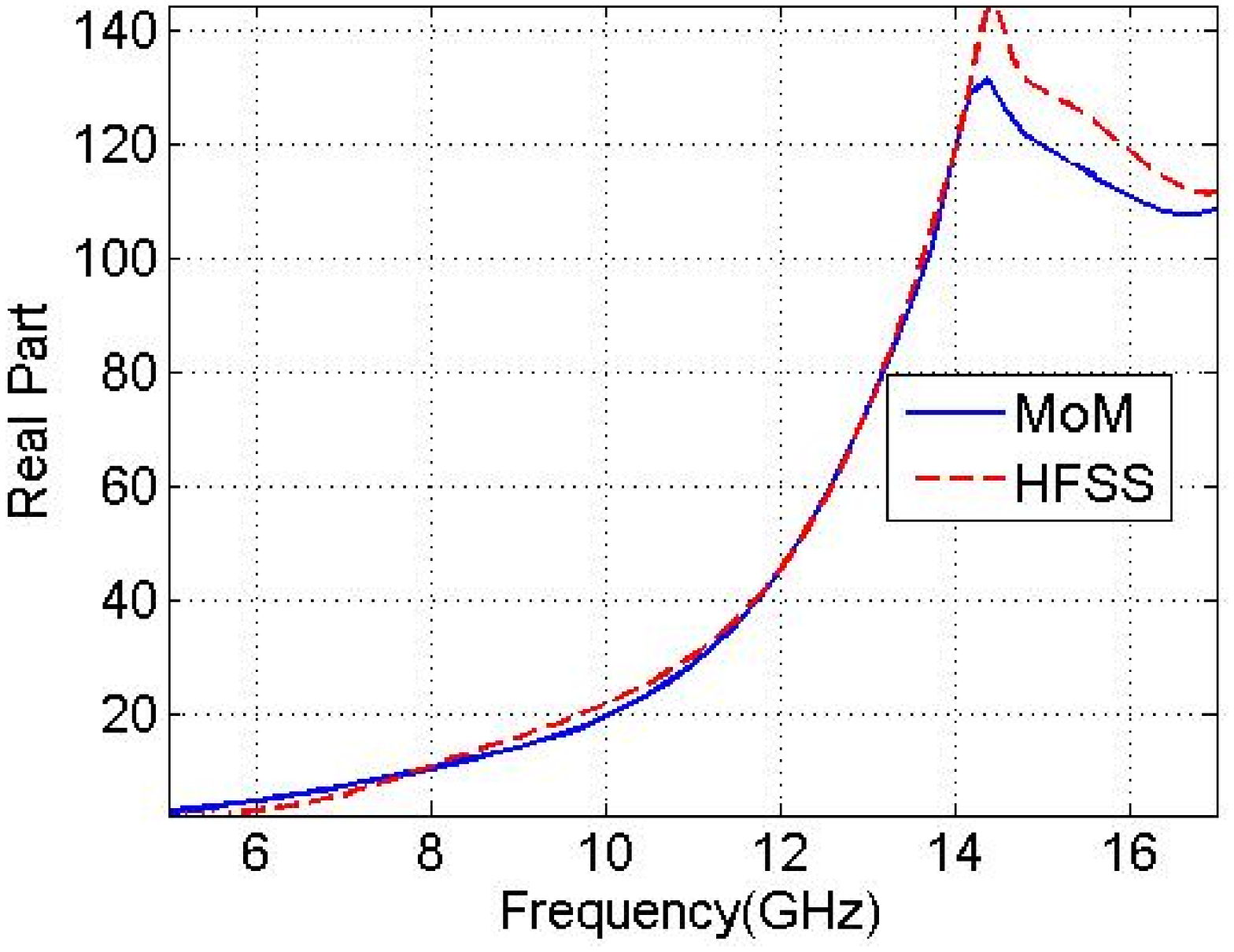}}}
      \hspace{0.2in}
      \subfigure{\scalebox{0.33}{\epsfig{file=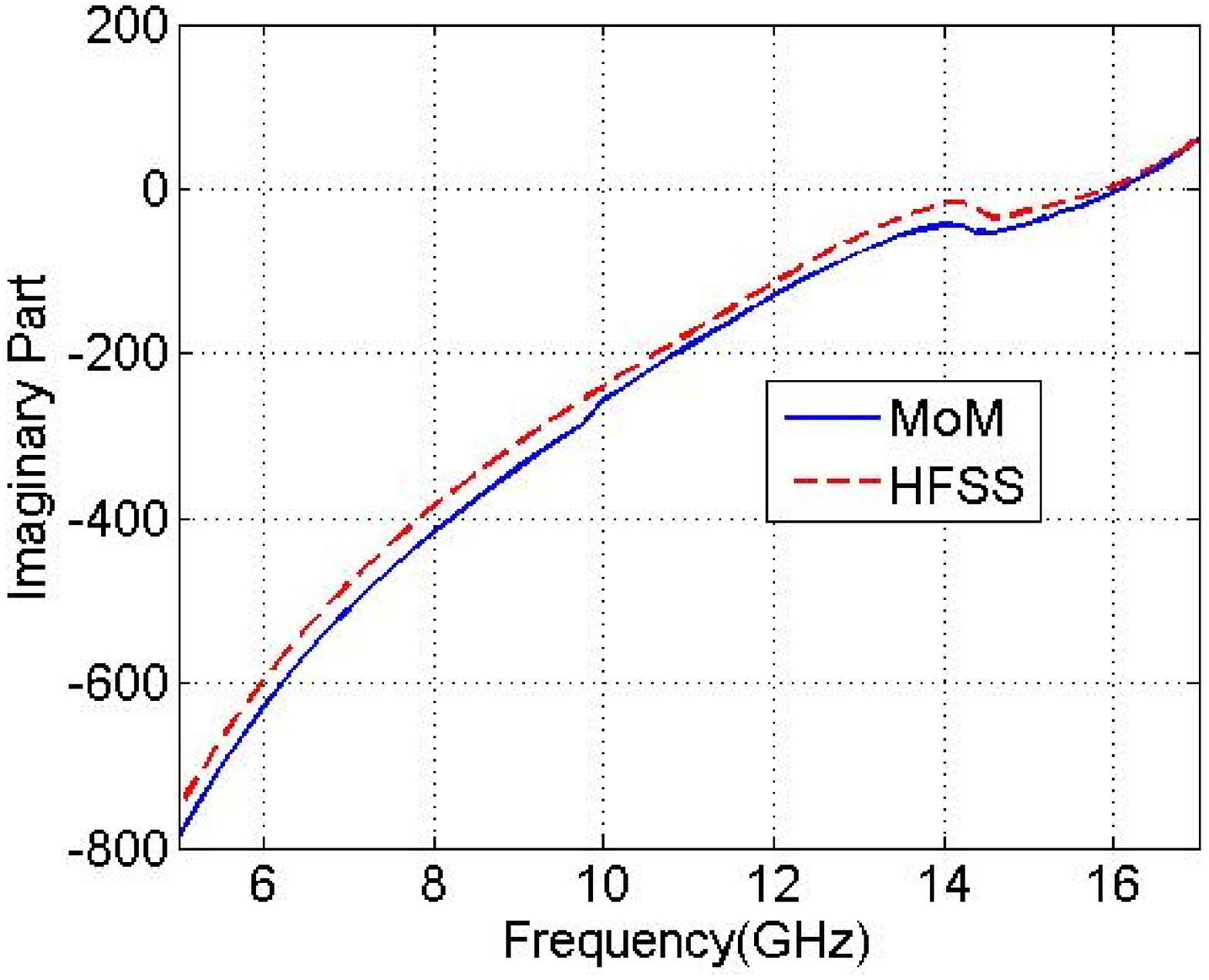}}}
  \caption{The input resistance and reactance. The antenna composed of a dipole of length $2l=9mm$, located at the center of a spherical core-shell resonator. Where $a_1=5.5~mm$,$a_2=7.5~mm$, $\epsilon_{r1}=10$ and $\mu_{r2}=2$.}
    \label{fig:compare_hfss}
\end{figure}

\quad Before we start the antenna designs, let us validate the accuracy of our technique. The geometry of the structure is shown in Fig.~\ref{fig:dipolecore}, where $a_1=5.5mm$ and $a_2=7.5mm$ and $l=4.5mm$. The material inside the core is a dielectric material described with $\epsilon_{r1}=10$ where the shell is a magnetic material (only permeability) with $\mu_{r2}=2$. The input impedance attained by MoM (Method of Moments) and HFSS is plotted in Fig.~\ref{fig:compare_hfss}. A very good comparison is observed validating our MoM program.

\quad It is worth noticing that the electric field inside the core is proportional to $n(n+1)P_n(\cos\theta\cos\theta')R_{12}\hat{J}_n(k_1r')\hat{J}_n(k_1r)$. For computation of the input impedance $\mathbf{r}$ and $\mathbf{r'}$ varies over the dipole, therefore $|\cos\theta\cos\theta'|=1$ and $r,r'<l$ . Thus, each term is basically proportional to $n(n+1)R_{12}\hat{J}_n(k_1r')\hat{J}_n(k_1r)$ where $r,r'<l$. So let us introduce a new term called `scattering term' for better perception. The scattering term is defined as:
\begin{equation}
\label{eq:scatterm}
ST = n(n+1)R_{12}\hat{J}^2_n(k_1l)
\end{equation}
For a small antenna the resonant behavior arise from the discontinuity of the sphere rather than the dipole. So investigating a scattering term such as \eqref{eq:scatterm} in addition to the input impedance seems appropriate to provide a physical insight of the electric field and resonant of the radiator.

\quad In the next section we discuss the quality factor behaviors of different cases for dipole antenna embedded inside a spherical resonator. We illustrate that by using magnetic shell and dielectric core, the quality factor of the antenna can approach the Chu limit.

\section{Low-Q Electrically Small Antennas }
\quad In this section we discuss different methods of antenna miniaturization in microwave frequencies.
As mentioned earlier, the simplest way to achieve antenna miniaturization is to use a high permittivity material. However, because of the strong concentration of electromagnetic fields inside the material, the performance of antenna is significantly degraded which brings the narrow-band characteristic. Recently there have been some works to properly overcome this problem by either printing the antenna on a magneto-dielectric substrate \cite{hossein}, or embedding the antenna inside a magnetic material \cite{hossein,ref4}. Here, we apply the developed formulations to present a better physical insight of these approaches. Moreover, we illustrate a novel design for lowering the Q by embedding the antenna inside a spherical core-shell where core is made of high dielectric and shell is made of magnetic materials.

\quad As highlighted in \cite{hossein,ref4}, by using a magneto-dielectric material one can deduce the capacitive behavior of the high dielectric resonator. Basically, by choosing moderate values for $\epsilon _{r} $ and $\mu _{r} $ the same miniaturization factor as a high permittivity non-magnetic material can be achieved, while the capacitive property of the resonator is now decreased by addition of some inductance behavior caused by $\mu _{r} $. This will allow us to improve the antenna impedance bandwidth. It has to be mentioned that in this case a higher input resistance is appeared since the impedance is related to the ratio of ${\mu _{r}\mathord{\left/{\vphantom{\mu _{r}\epsilon_{r}}}\right.\kern-\nulldelimiterspace} \epsilon _{r}}$. By increasing the $\mu_r$ the input resistance goes higher and the resonance occurs at lower frequencies, hence
the same miniaturization (as the high $\epsilon_r$ situation) is achieved even with smaller permeabilities. This shows that by embedding a dipole antenna inside a magnetic material lower quality factor can be achieved however, the quality factor is still much higher than the $Q_{Chu}$.
Another approach to achieve a low Q antenna is to embed the antenna inside an ENG material which will also be illustrated in this paper.

\quad Here, we design a spherical core-shell, where the core is filled with high-dielectric materials and the shell is made of large magnetics materials. We show that, if the shell is thin and filled with a material with a large $\mu_r$, the field inside the core is negligible compared to the field of the dipole and the electric field inside the shell. This can be interpreted as almost zero absorbed energy inside the core. Also, since the shell is thin the absorbed energy inside the shell is small compared to the energy absorbed outside the resonator. Thereafter, one can expect to achieve a quality factor close to Chu limit. Before we begin, let us first take a look at the quality factor.

Assume that, the antenna is tuned at a frequency $\omega_0$ with a series reactance $X_s(\omega)$. Thus, the total reactance is zero at $\omega_0$. if the input impedance of the tuned antenna is $Z_0(\omega)$, where the input impedance of the antenna is $Z(\omega)$ we have:
\begin{subequations}
\begin{align}
&X_0(\omega_0)=X(\omega_0)+X_s(\omega_0)=0\\
&Z_0(\omega)=Z(\omega)+jX_s(\omega)=R_0(\omega)+jX_0(\omega)
\end{align}
\end{subequations}
Where $R_0$ and $X_0$ are the resistance and reactance of the tuned antenna respectively. If the antenna is matched to $R_0(\omega_0)$, the reflection would be zero at $\omega_0$ provided that the tuning inductance or capacitance is lossless. For a tuned antenna at $\omega=\omega_0$ the matched VSWR bandwidth can be well-defined for all frequencies.
The bandwidth $\Delta\omega$ is defined as the difference between the two frequencies on other side of $\omega_0$ at which the reflection equals to a constant number $\alpha$, with $\alpha<1/2$. If $|\Delta\omega/\omega_0|\ll 1$, the solution for bandwidth is obtained as \cite{yaghiq}:
\begin{align}
\label{eq:deltaw}
&\Delta\omega=2\sqrt{\frac{\alpha}{1-\alpha}}~~\frac{R_0(\omega_0)}{Z'(\omega_0)}\\
&\alpha=\frac{|R_0(\omega_\pm)-R_0(\omega_0)|^2+|X_0(\omega_\pm)|^2}{|R_0(\omega_\pm)+R_0(\omega_0)|^2+|X_0(\omega_\pm)|^2}
\end{align}
It is renowned that the quality factor is inversely proportional to the bandwidths. Hence the larger the bandwidths the smaller the quality factor and vice versa. To verify this theory, let us obtain the quality factor. Yaghjian and Best shown that the quality factor for an antenna tuned to zero reactance at  $omega=\omega_0$ can be defined as \cite{yaghiq}:
\begin{align}
\label{eq:q2}
&Q(\omega_0)=\left|\frac{\omega_0}{2R_0(\omega_0)}X'_0(\omega_0)-\frac{2\omega_0}{|I_0|^2R_0(\omega_0)}[W_L(\omega_0)+W_R(\omega_0)]\right|
\end{align}
where $W_L(\omega_0)+W_R(\omega_0)$ is the total dispersion energy. Using the RLC circuit realization, the quality factor can be well approximated by \cite{yaghiq}:
\begin{equation}
\label{eq:q1}
Q(\omega_0)\approx\frac{\omega_0}{2R_0(\omega_0)}|Z'_0(\omega_0)|
\end{equation}
According to the above equation, the quality factor is directly related to the input impedance and inversely proportional to Bandwidths. This means that by having the input impedance of any antenna we can easily obtain the Q.

\begin{figure}[htbp]
  \centering
           \subfigure[]{\scalebox{0.45}{\epsfig{file=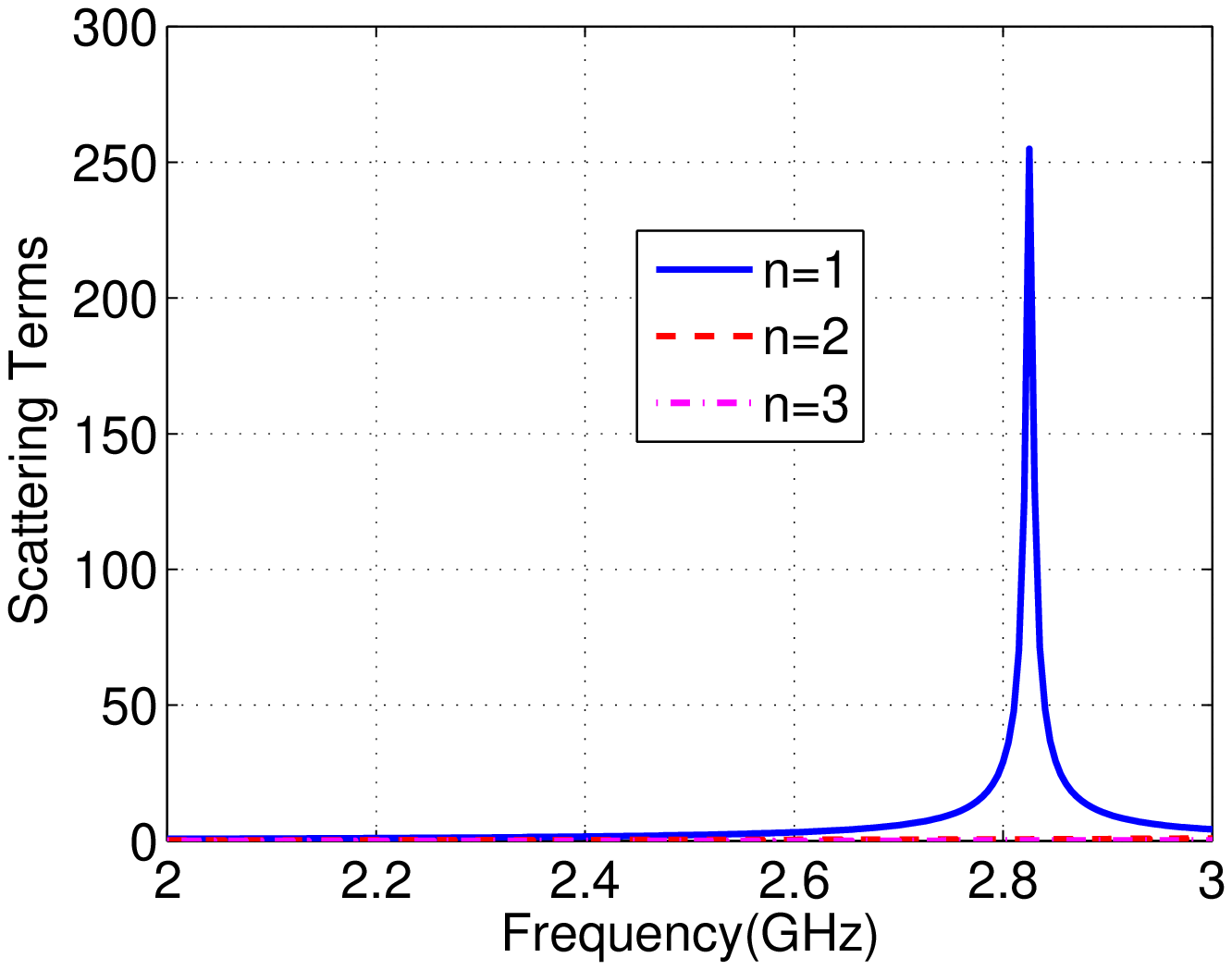}}}
           \hspace{0.2in}
           \subfigure[]{\scalebox{0.45}{\epsfig{file=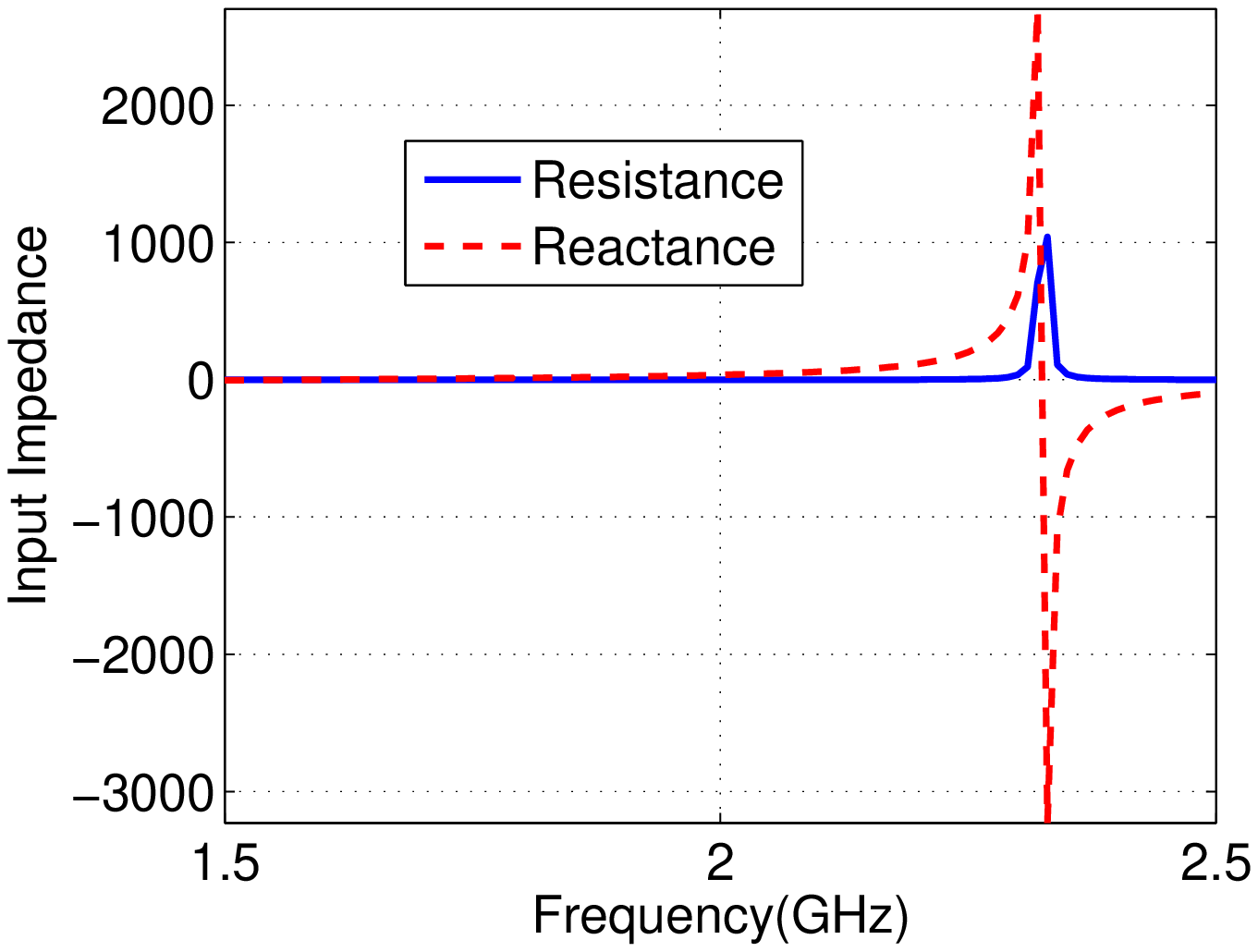}}}
  \caption{ (a) Scattering term for the first three modes. (b) The input resistance and reactance. The antenna composed of a dipole with total length of $2l=9mm$, located at the center of a spherical resonator filled with high dielectric material of $\epsilon_{r1}=100$. The radius of the sphere is $a_1=7.5~mm$.}
    \label{fig:highd1}
\end{figure}

\begin{figure}[htbp]
  \centering
           \subfigure[]{\scalebox{0.45}{\epsfig{file=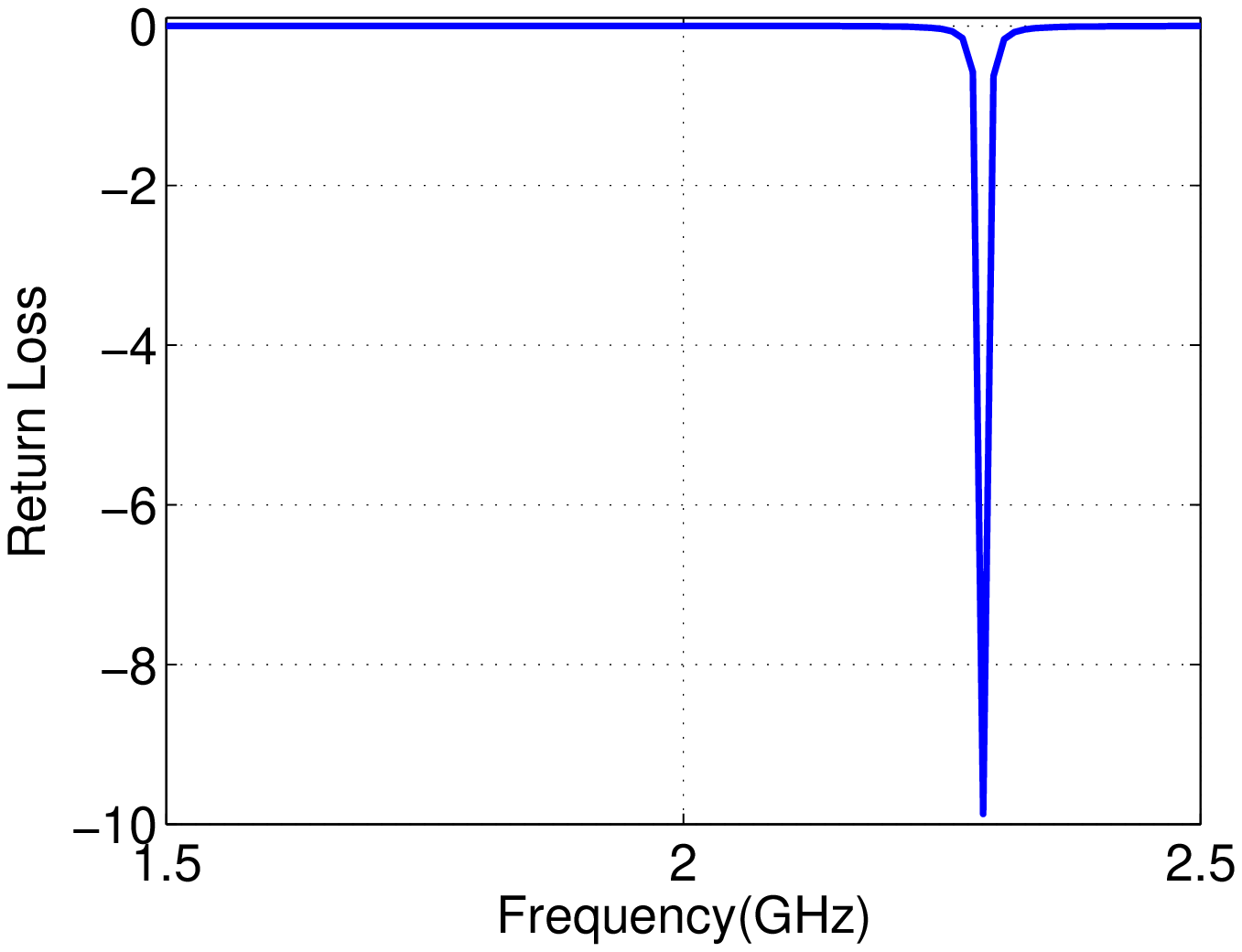}}}
  \caption{The return loss of the antenna in Fig.~\ref{fig:highd1}. The return loss presents a very narrow-band bandwidths which is consistent with the Q=293.1.}
    \label{fig:return_highd1}
\end{figure}

\quad As discussed earlier to achieve the desired miniaturization, the simplest way is to use high-dielectric material. The objective it to have a wide-band antenna working around 2GHz. We embed a dipole antenna at the center of a  dielectric sphere filled with a high dielectric material . The spherical radiator has a radius of $a_1=7.5~mm$ and is filled with dielectric material with $\epsilon_{r1}=100$. It is worth mentioning that, a spherical resonator is basically a core-shell where the material of the shell is free space. The scattering terms (for the first three modes) and input impedance are shown in Fig.~\ref{fig:highd1}. As is shown, the first scattering term  behavior exhibits a resonance around $2.8GHz$, so we expect to have a resonance in the input impedance around this region. The input impedance shows the resonance performance around $2.3GHz$. The return loss for the tuned antenna is plotted in Fig.~\ref{fig:return_highd1}(a), revealing  a narrow-band resonance at $f=2.32~GHz$.
For this resonance, the quality factor of the tuned antenna calculated by using \eqref{eq:q1} is $Q=293.1$, whereas the Chu limit is $Q_{chu}=\frac{1}{ka^3}+\frac{1}{ka}=23.69$.
The obtained Q from the antenna is more than 12 times the Chu limit which was expected from the return loss behavior.

\begin{figure}[htbp]
  \centering
      \subfigure[]{\scalebox{0.45}{\epsfig{file=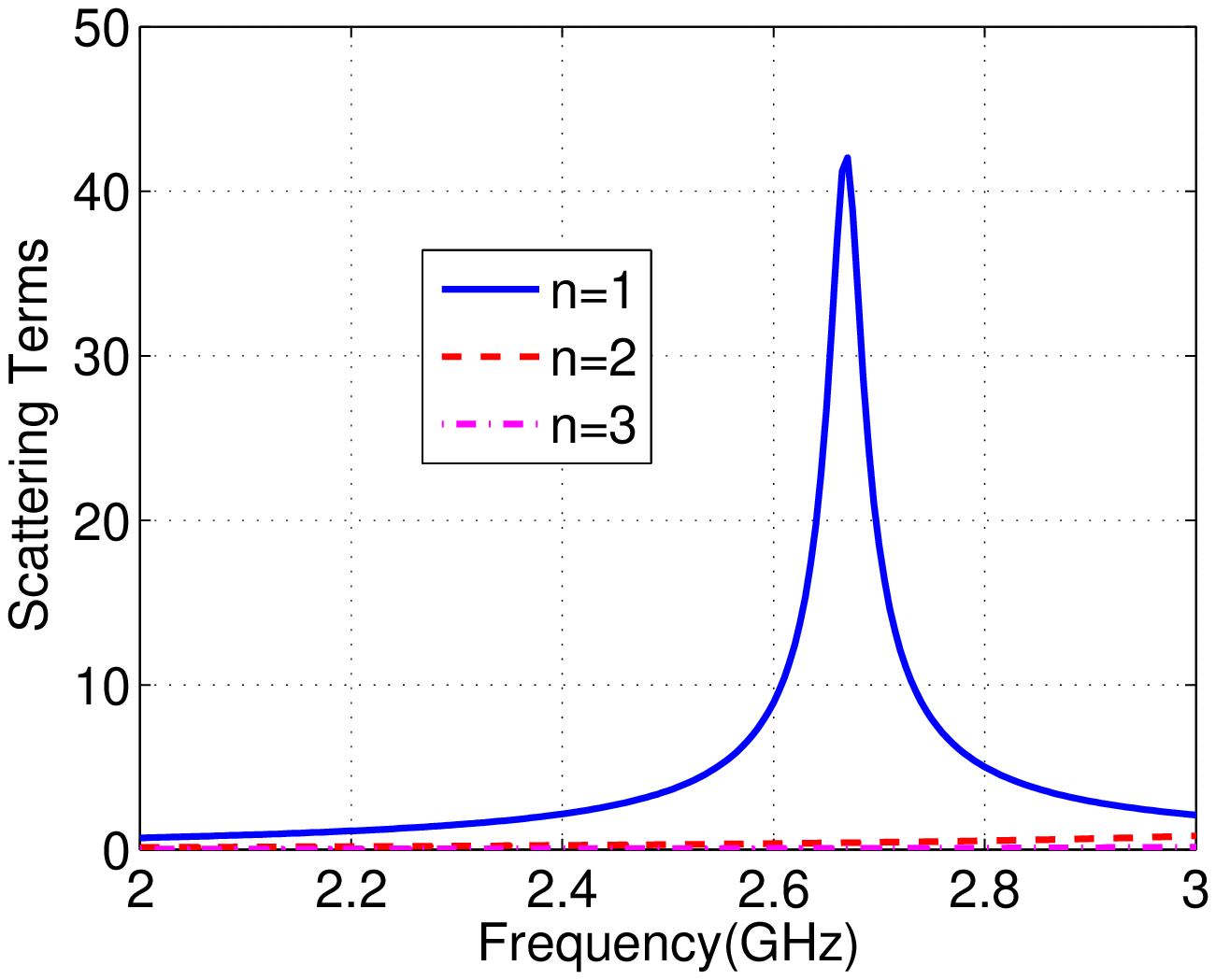}}}
      \subfigure[]{\scalebox{0.45}{\epsfig{file=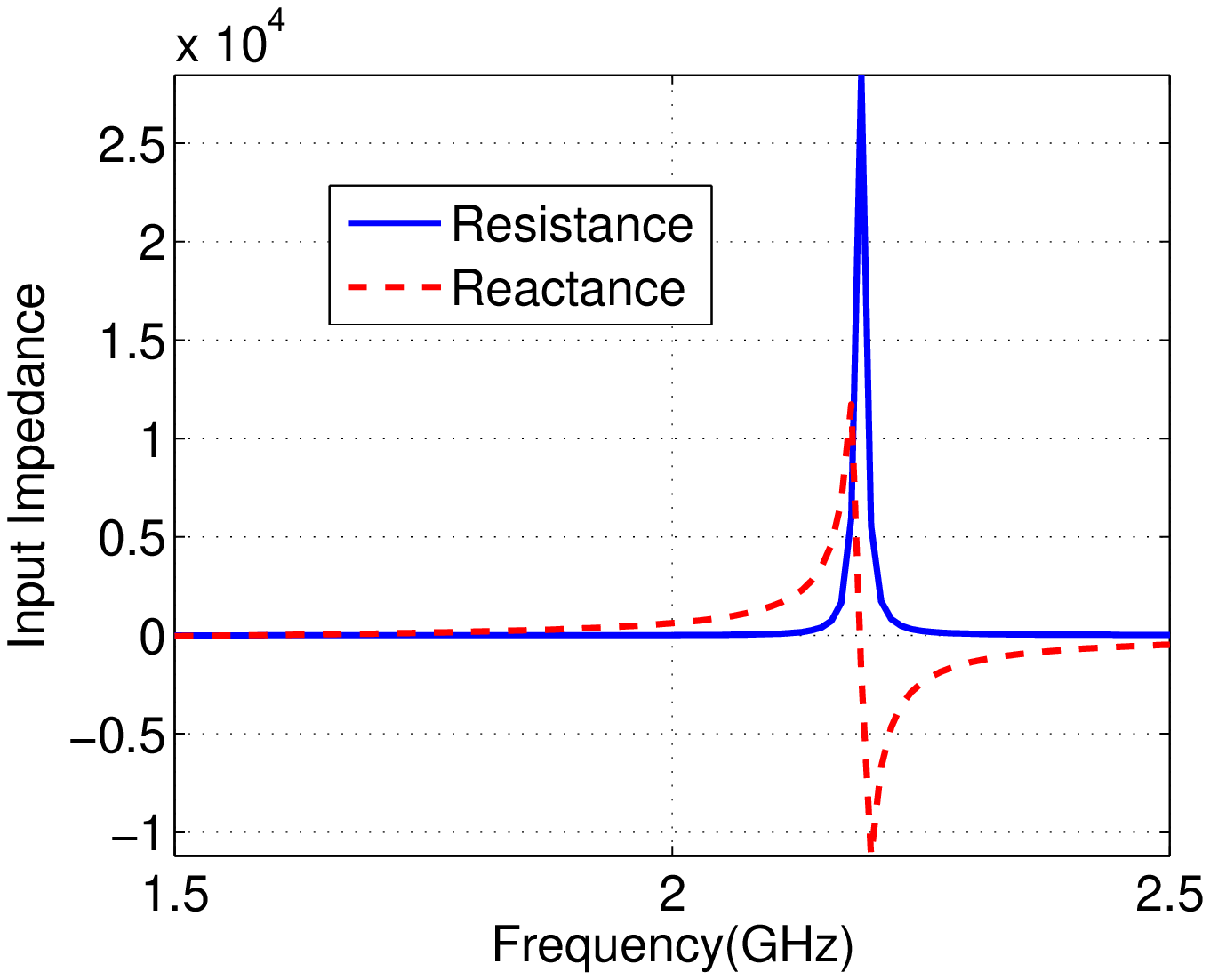}}}
  \caption{(a) Scattering term for the first three modes. (b) The input resistance and reactance. The antenna composed of a dipole with total length of $2l=9mm$, located at the center of a spherical resonator filled with magneto-dielectric material of $\epsilon_{r1}=12$ and $\mu_{r1}=8$. The radius of the sphere is $a_1=7.5~mm$.}
    \label{fig:mu_eps1}
\end{figure}

\begin{figure}[htbp]
  \centering,
           \subfigure[]{\scalebox{0.45}{\epsfig{file=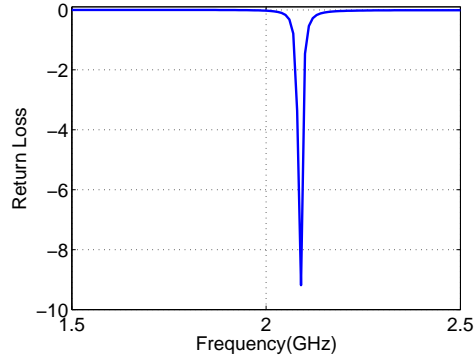}}}
  \caption{The return loss of the antenna in Fig.~\ref{fig:mu_eps1}. The return loss presents a narrow bandwidths which is consistent with the Q=101.3.}
    \label{fig:return_mueps}
\end{figure}

\quad The input reactance in Fig.~\ref{fig:highd1}(b) demonstrates a very high capacitive behavior.
To increase the bandwidths and hence reduce the quality factor, we can introduce an artificial inductance. The artificial inductor caused by permeability reduces the capacitance characteristics of the input impedance and hence a wider bandwidths can be achieved. To verify this theory, instead of embedding the antenna inside a high-dielectric material, one can utilize magneto-dielectric materials. Fig.~\ref{fig:mu_eps1} demonstrates the scattering terms and the input impedance for such an antenna embedded inside a magneto-dielectric material ($\epsilon_r=12,\mu_r=8$). Notice that, the first scattering term, of particular interest, shows a resonance performance with a higher bandwidths compared with the one depicted in Fig.~\ref{fig:highd1}(a). The return loss is shown in Fig.~\ref{fig:return_mueps} exhibiting a resonant around $f=2.19~GHz$. Clearly the bandwidths has increased, meaning the quality factor should be reduced. From \eqref{eq:q1}, quality factor is equal to $Q=101.3$ which is 3.64 times the $Q_{chu}=27.82$. This result verifies that by adding magnetic material Q has decreased while bandwidths has increased.

\begin{figure}[htbp]
  \centering
      \subfigure[]{\scalebox{0.45}{\epsfig{file=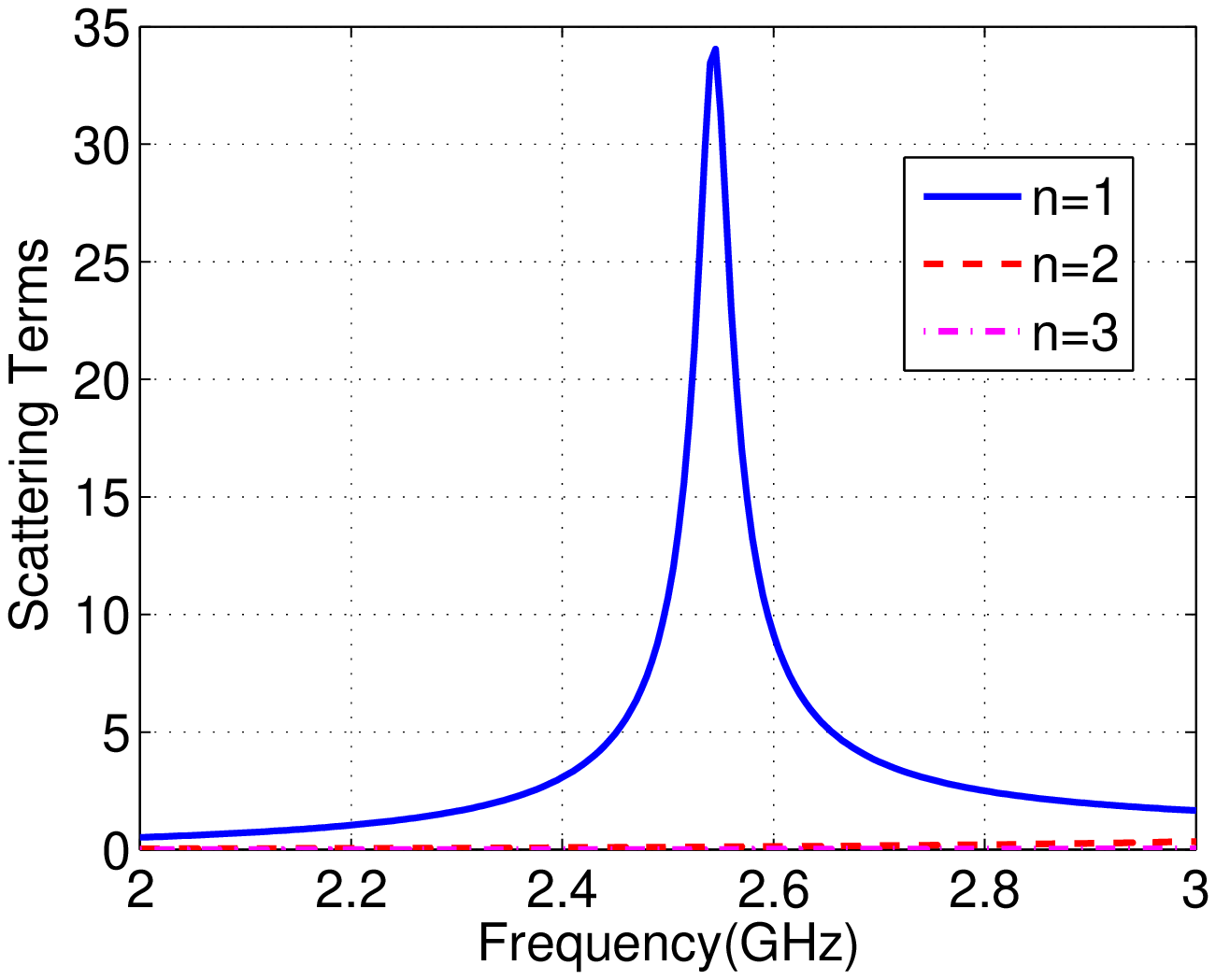}}}
      \subfigure[]{\scalebox{0.45}{\epsfig{file=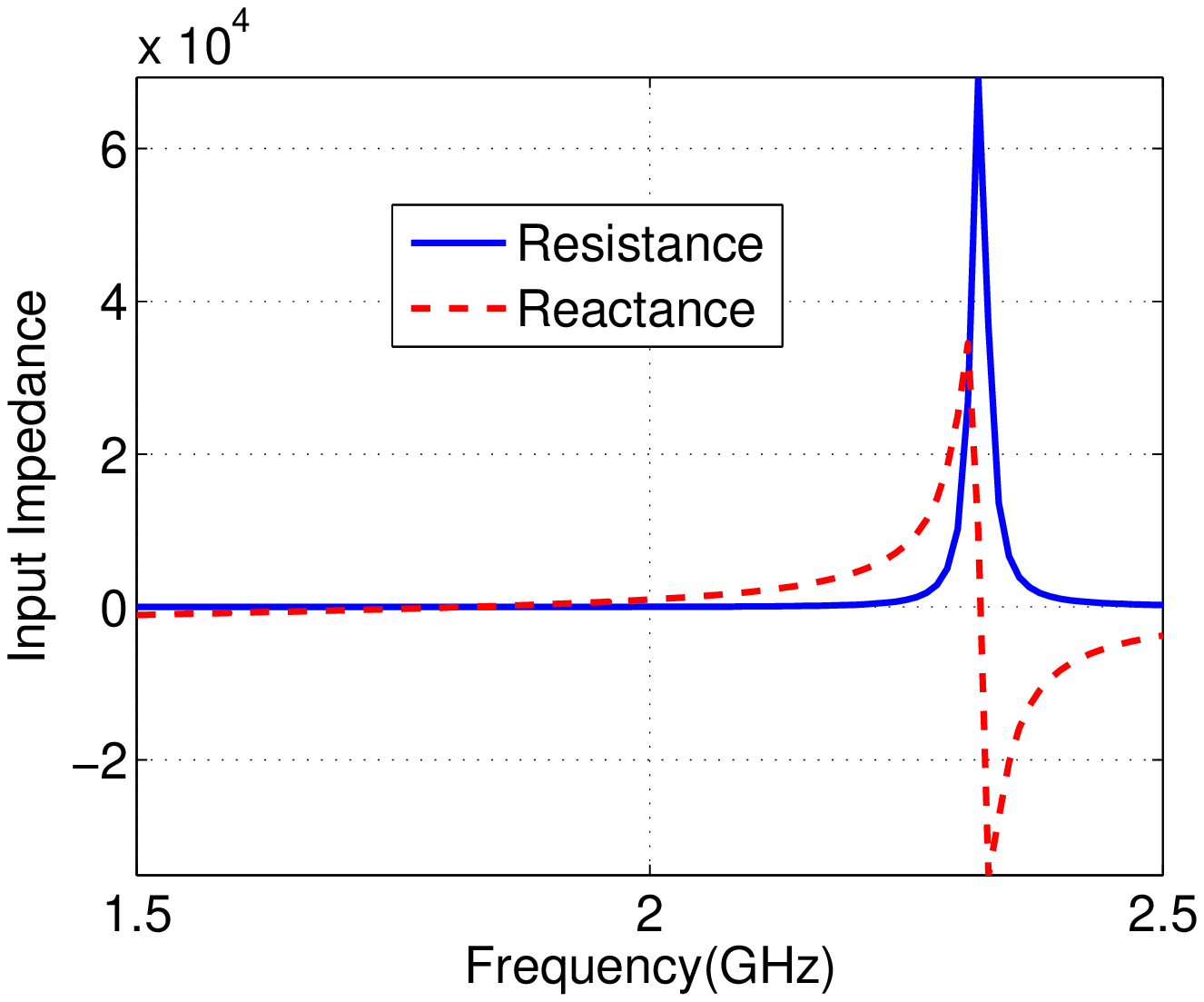}}}
  \caption{(a) Scattering term for the first three modes. (b) The input resistance and reactance. The antenna composed of a dipole with total length of $2l=9mm$, located at the center of a spherical resonator filled with magnetic material of $\mu_{r1}=60$. The radius of the sphere is $a_1=7.5~mm$.}
    \label{fig:mu1}
\end{figure}

\begin{figure}[htbp]
  \centering,
           \subfigure[]{\scalebox{0.45}{\epsfig{file=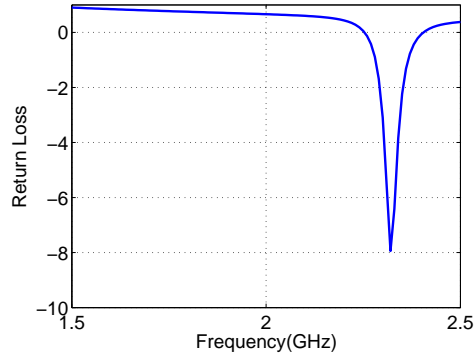}}}
  \caption{The return loss of the antenna in Fig.~\ref{fig:mu1}. The return loss presents a larger bandwidths (compared to the previous case) which is consistent with the Q=81.89.}
    \label{fig:return_mu}
\end{figure}

\quad As shown earlier, by inserting artificial inductance the Q of the antenna reduces resulting in a better bandwidths performance. Hence, one can (in theory) embed the antenna inside a pure magnetic material to achieve a higher bandwidths while maintaining the same miniaturization.
Fig.~\ref{fig:mu1} demonstrates the scattering terms and the input impedance for a dipole antenna embedded inside a magnetic material with $\mu_r=60$. The return loss is shown in Fig.~\ref{fig:return_mu} exhibiting a resonant around $f=2.32~GHz$. Clearly the bandwidths has increased, resulting in a lower quality factor. From \eqref{eq:q1}, quality factor is equal to $Q=81.89=3.4Q_{Chu}$. This result verifies that by adding magnetic material, the capacitance performance of the input impedance decreases leading to a better radiation performance. Namely, the quality factor, i.e. the stored energy per radiated power diminished while matched bandwidths enhances.

\begin{figure}[htbp]
  \centering,
           \subfigure[]{\scalebox{0.45}{\epsfig{file=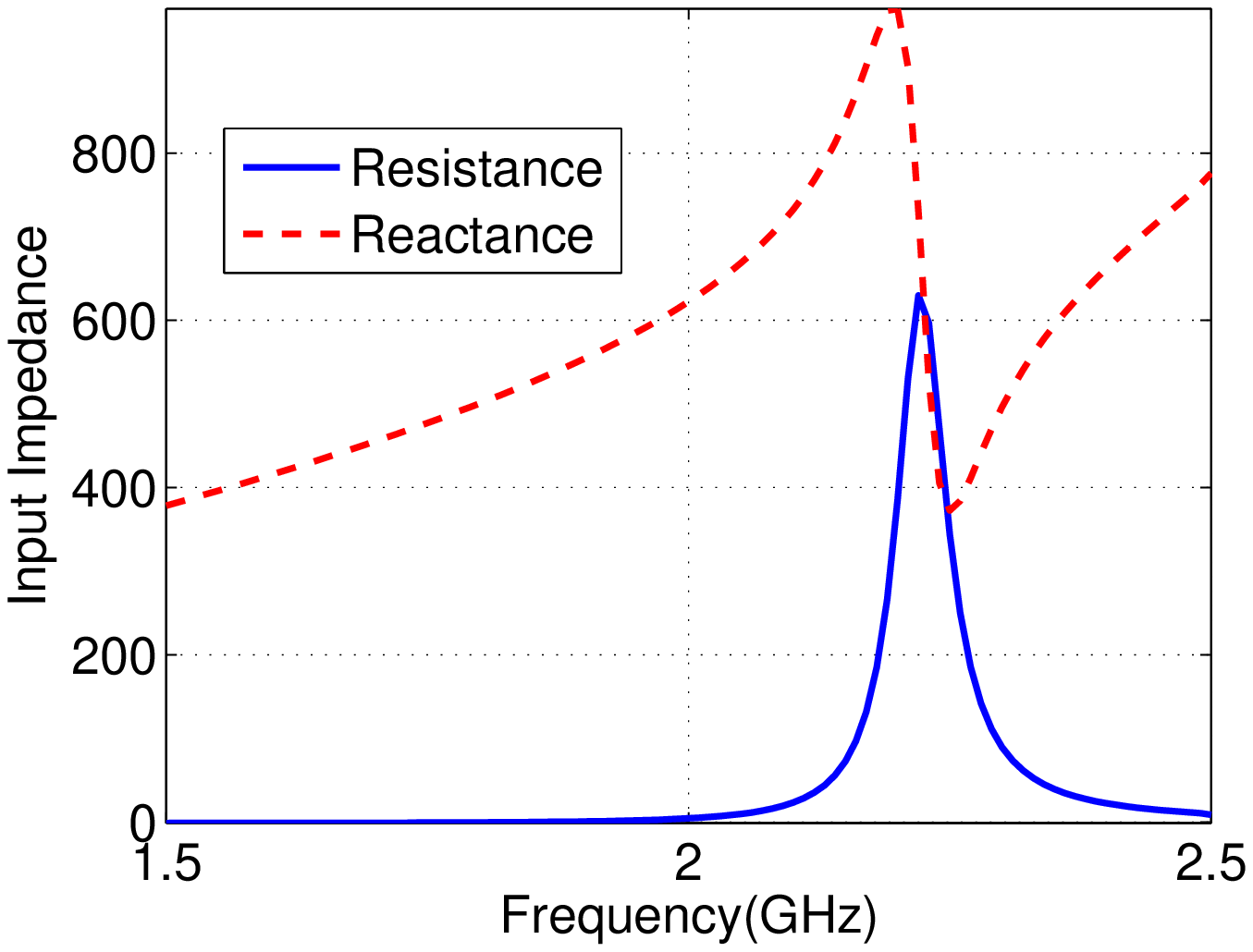}}}
           \subfigure[]{\scalebox{0.45}{\epsfig{file=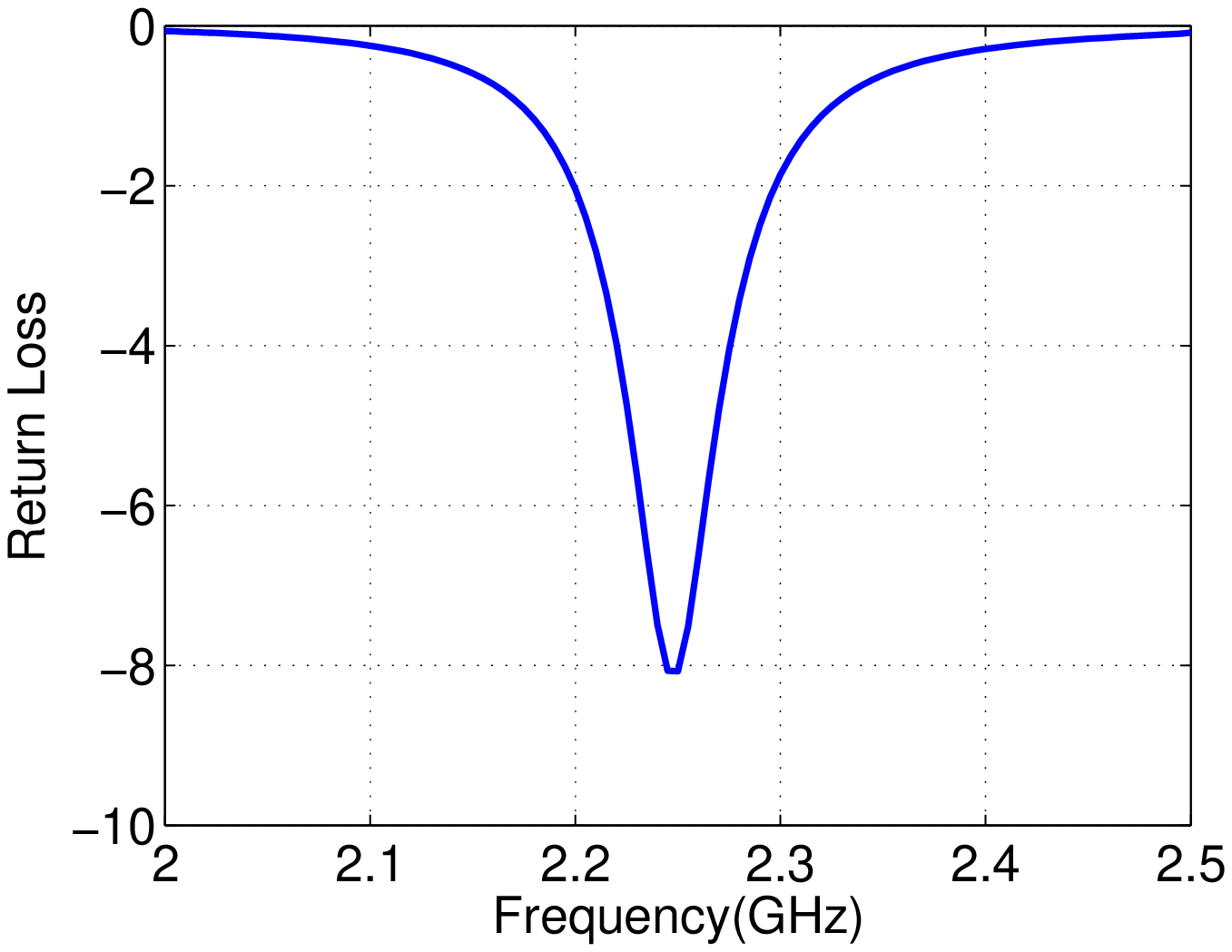}}}
           \caption{ (a) The input impedance and (b) return loss. The antenna composed of a dipole with total length of $2l=9mm$, located at the center of a spherical resonator filled with Drude material where $\epsilon_r=1-\frac{\omega_e^2}{\omega*(\omega+j\gamma e)}$ and $\omega_e=2 GHz,\gamma_e=0.001\omega_es$.}
    \label{fig:return_eng}
\end{figure}

\begin{figure}[htbp]
  \centering
      \subfigure[]{\scalebox{0.45}{\epsfig{file=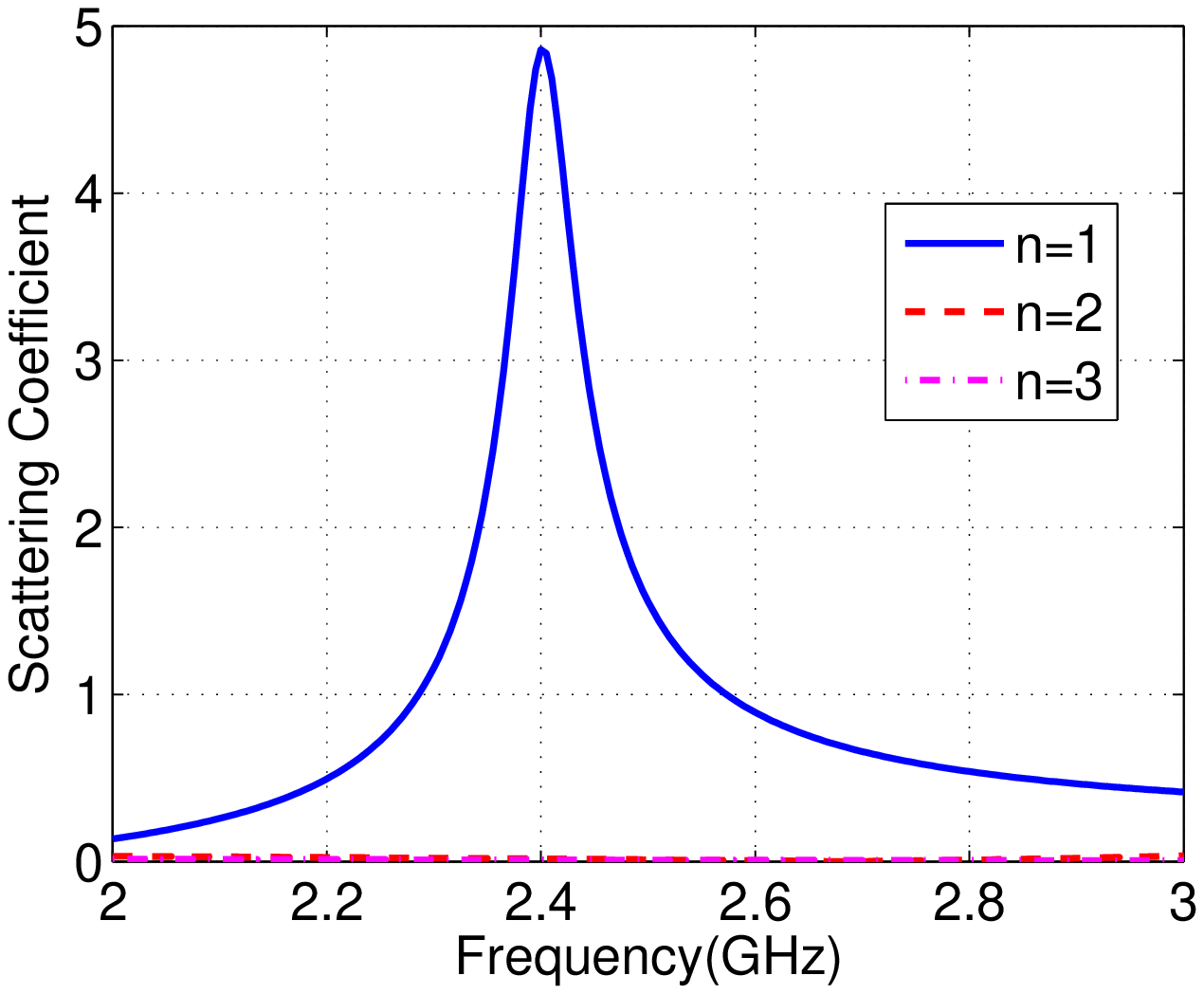}}}
      \subfigure[]{\scalebox{0.45}{\epsfig{file=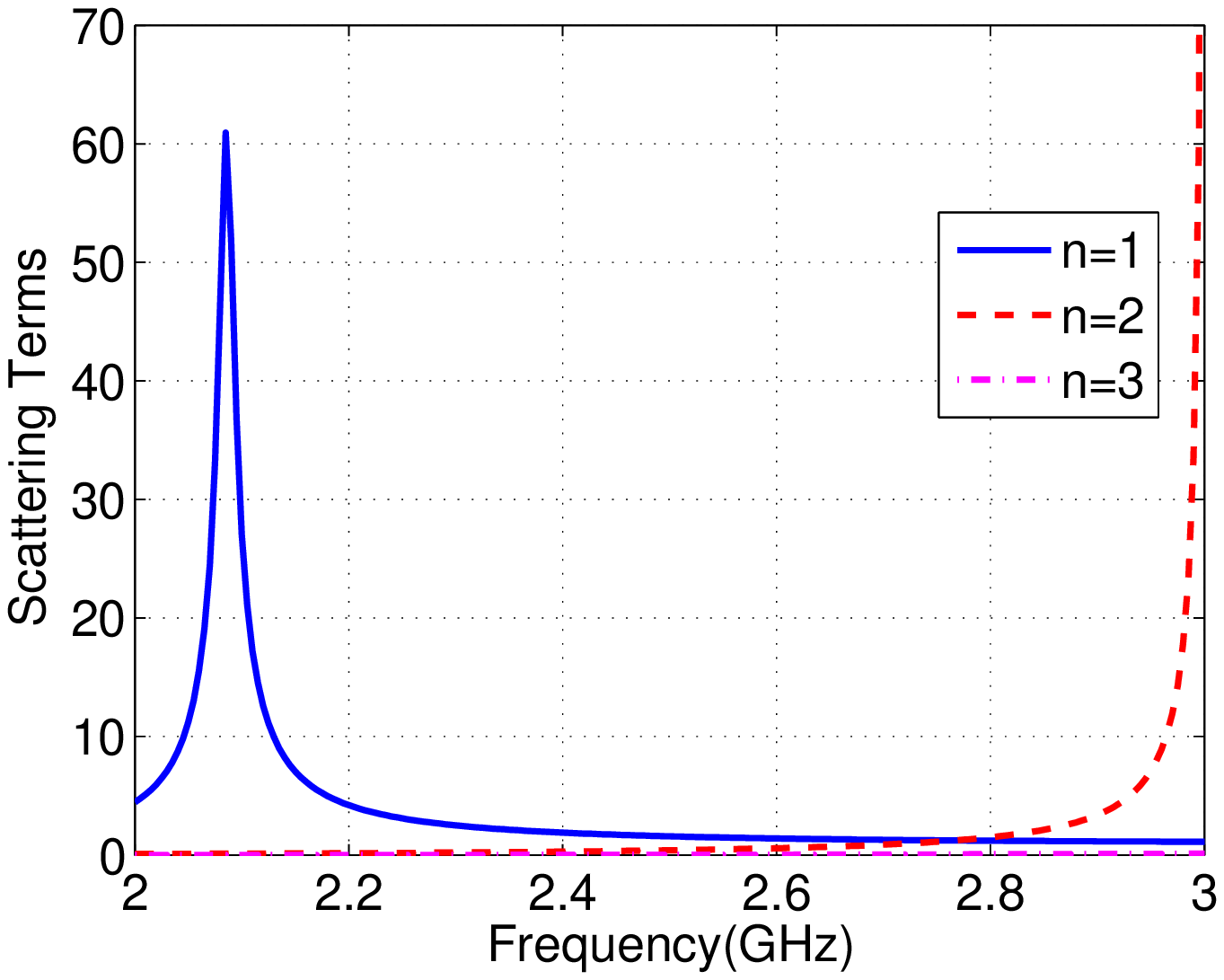}}}\\
      \subfigure[]{\scalebox{0.45}{\epsfig{file=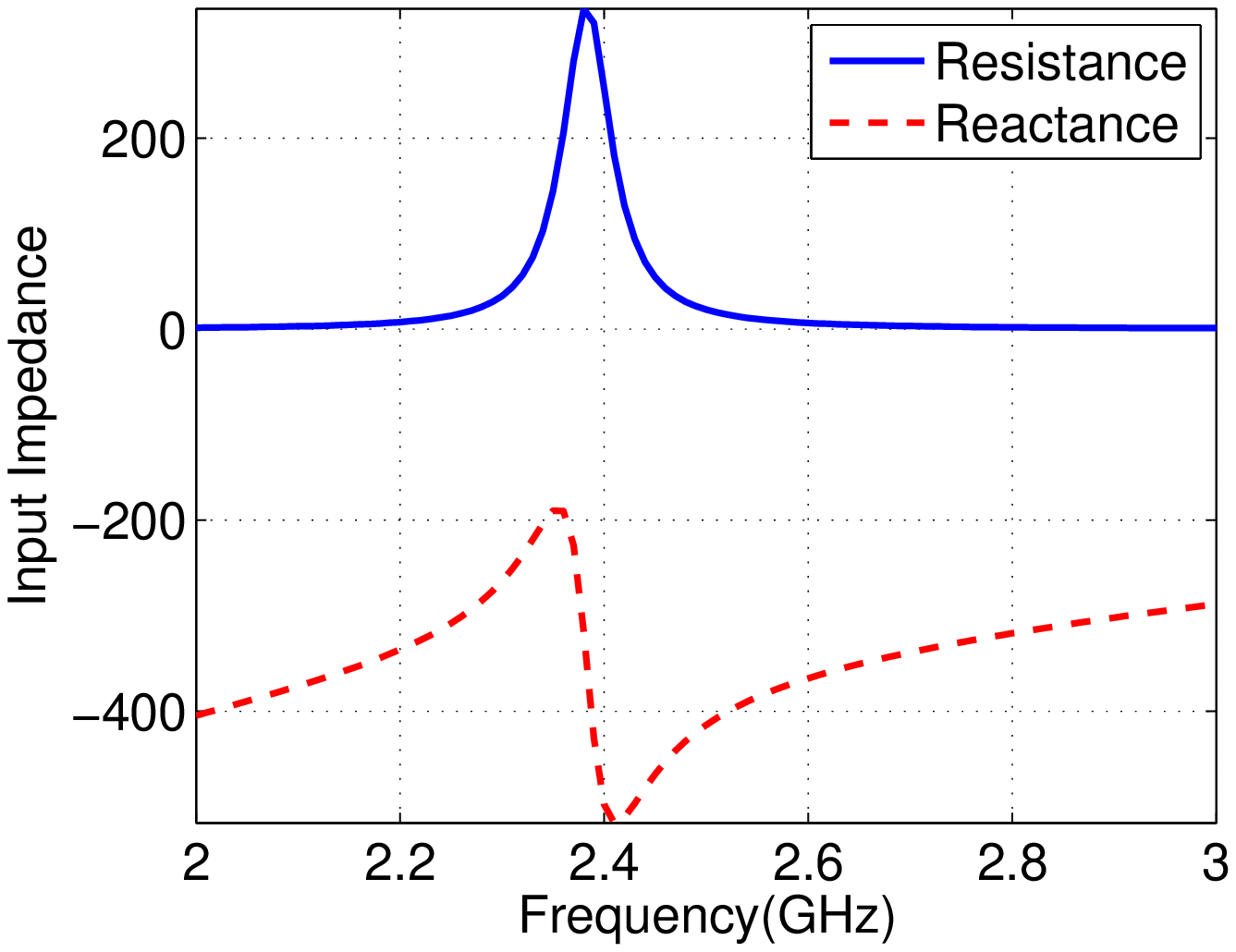}}}
  \caption{(a) Scattering terms for the first three modes, (b) Scattering terms inside the shell for the first three modes and, (c) The input resistance and reactance. The antenna composed of a dipole of length $2l=9mm$, located at the center of a spherical core-shell resonator. Where $a_1=6.5~mm$,$a_2=7.5~mm$, $\epsilon_{r1}=4$ and $\mu_{r2}=90$.}
    \label{fig:shell1}
\end{figure}

\begin{figure}[htbp]
  \centering,
           \subfigure[]{\scalebox{0.5}{\epsfig{file=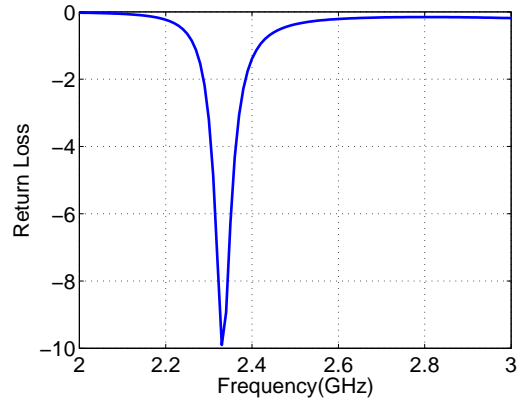}}}
  \caption{The return loss of the antenna in Fig.~\ref{fig:shell1}. The return loss presents a large-bandwidth which is consistent with the Q=34.97.}
    \label{fig:return_shell1}
\end{figure}

\quad Although, embedding the antenna inside magneto-dielectric or magnetic materials increases the bandwidths of the antenna, but the quality factor as low as Chu limit is still not accessible. It has be demonstrated in \cite{stuart} that by embedding the antenna inside an ENG material, a quality factor as low as $1.5Q_{Chu}$ can be achieved. A demonstration has been shown  (cf. \ref{fig:return_eng}) in this work where we embed the antenna inside a negative permittivity and the antenna shows a quality factor around $Q=1.53Q_{Chu}$.
However, to the best of our knowledge there is no report of achieving the actual Chu limit. In this paper, we present a novel design to approach the Chu limit by embedding the antenna inside a magnetic shell.
The idea is to reduce the stored energy inside the radiator to get closer to the Chu limit. Hence, it seems reasonable to embed the antenna inside a spherical core-shell with a very thin shell. If the absorbed energy inside the core approaches zero compared to the field of the dipole
the achieved $Q_{Chu}$. Fig.~\ref{fig:shell1} demonstrates the scattering terms and the input impedance for a dipole of length $2l=9~mm$ embedded inside a core-shell. The inner and outer radii of the core are $a_1=6.5~mm$ and $a_2=7.5~mm$. The core is filled with a dielectric with permittivity of $\epsilon_{r1}=4$, while the shell is a magnetic material of $\mu_{r2}=90$. The scattering terms show a significant increase in the bandwidths of the resonance hence, we expect to have a higher bandwidths and lower Q for the antenna.
Also, note that the scattering coefficient in the shell is the same as the scattering coefficient inside a  sphere of radius $7.5~mm$ filled with $\mu_r=90$. As is evident form Fig.~\ref{fig:shell1}(a) and (b), the scattering terms inside the shell is much stronger than the ones in the core. Hence, it is anticipated that the field inside the shell to  be large. Nevertheless, the shell is thin and the total energy stored inside the resonator is still small compared to stored energy outside. The return loss is shown in Fig.~\ref{fig:return_shell1} exhibiting a resonant around $f=2.36~GHz$. Clearly the bandwidths is increased remarkably, meaning the quality factor should be reduced. From \eqref{eq:q1}, quality factor is equal to $Q=34.97=1.56Q_{Chu}$. This is an interesting results, exhibiting that by surrounding the antenna with a magnetic shell the quality factor can be reduced. By optimizing the structure reduced Q can be obtained as will discuss later.

\quad Since the scattering term can provide the resonance behavior along with the bandwidth performance of the antenna. Thus, let us first investigate the behavior of the scattering term with changing the material of the core.
\begin{figure}[htbp]
  \centering,
           \subfigure[]{\scalebox{0.45}{\epsfig{file=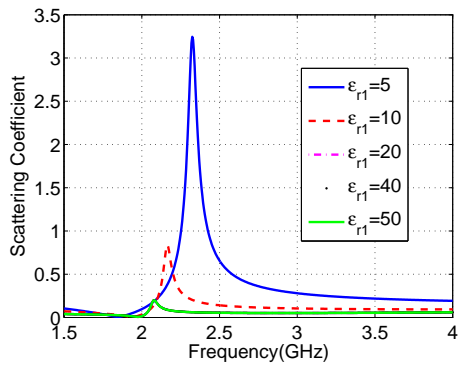}}}
           \subfigure[]{\scalebox{0.45}{\epsfig{file=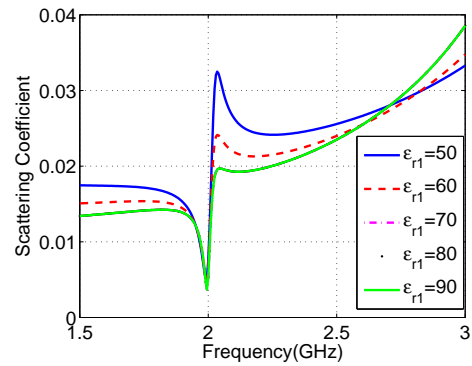}}}
  \caption{The scattering term vs. different dielectric material of the core. The core-shell is described as follows: $a_1=6.5~mm, a_2=7.5~mm$, $\mu_{r2}=90$. A dipole of length $2l=9~mm$ is located at the center of the core-shell.}
    \label{fig:scat_term}
\end{figure}

\begin{figure}[htbp]
  \centering
      \subfigure[]{\scalebox{0.45}{\epsfig{file=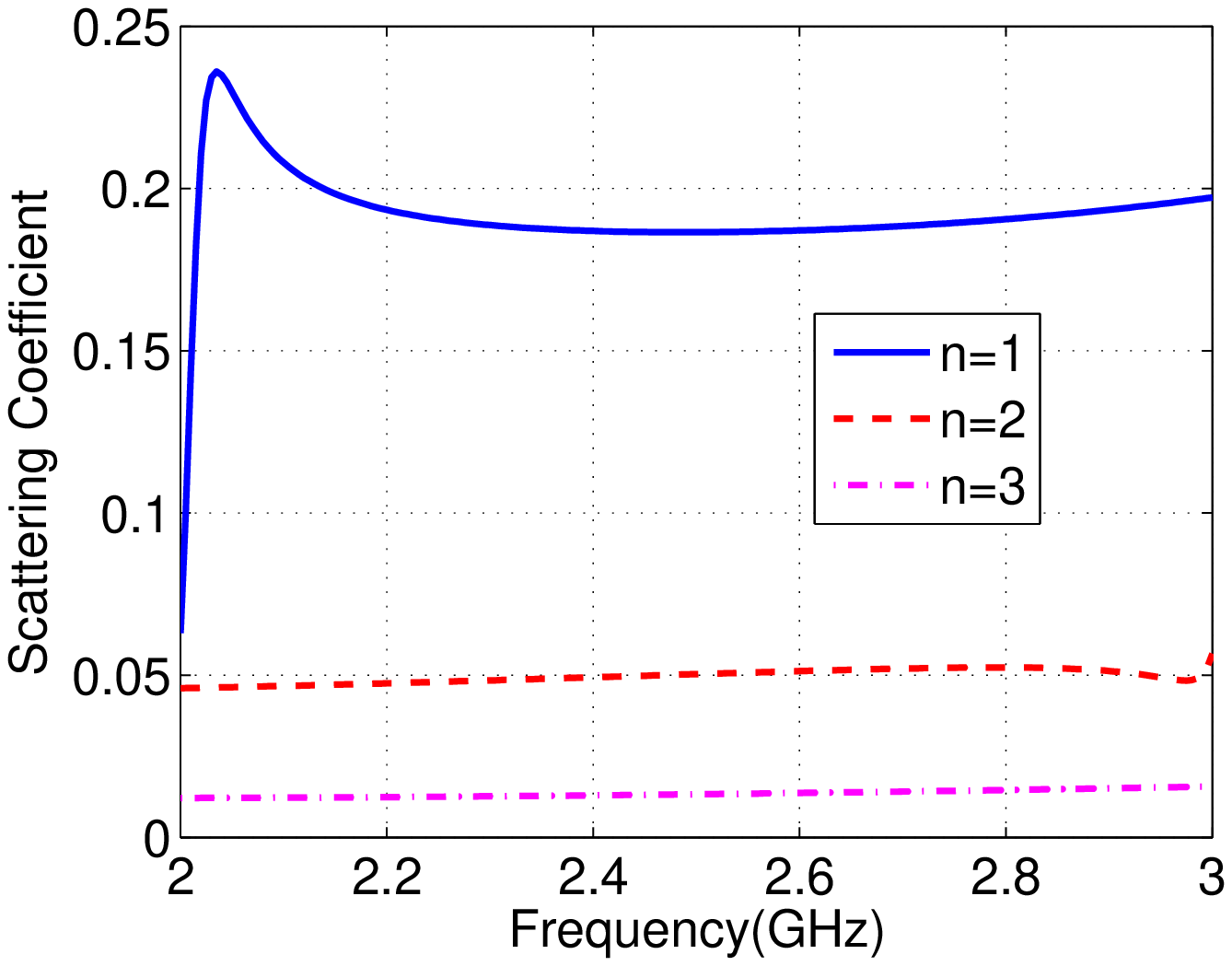}}}\\
      \subfigure[]{\scalebox{0.45}{\epsfig{file=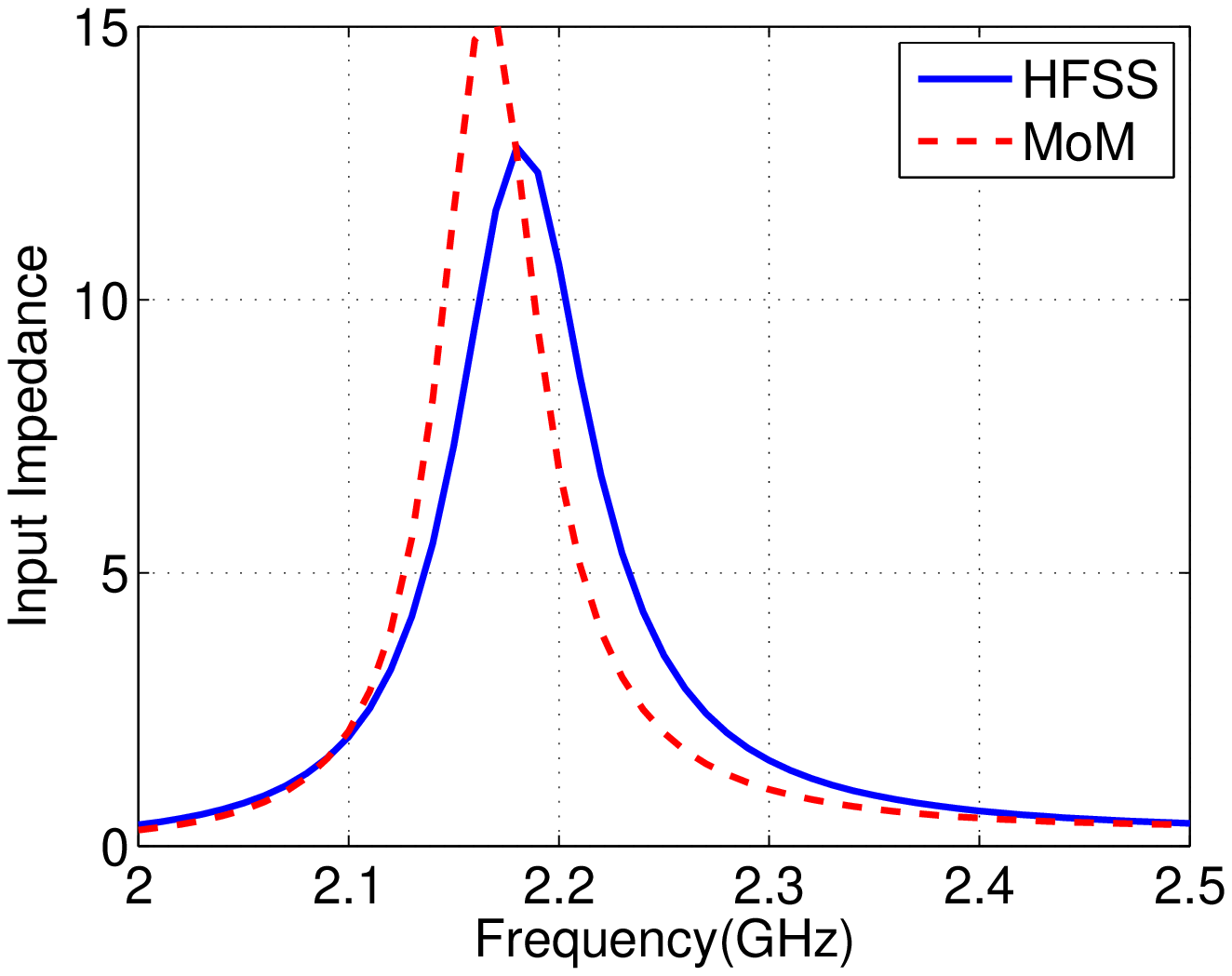}}}
      \subfigure[]{\scalebox{0.45}{\epsfig{file=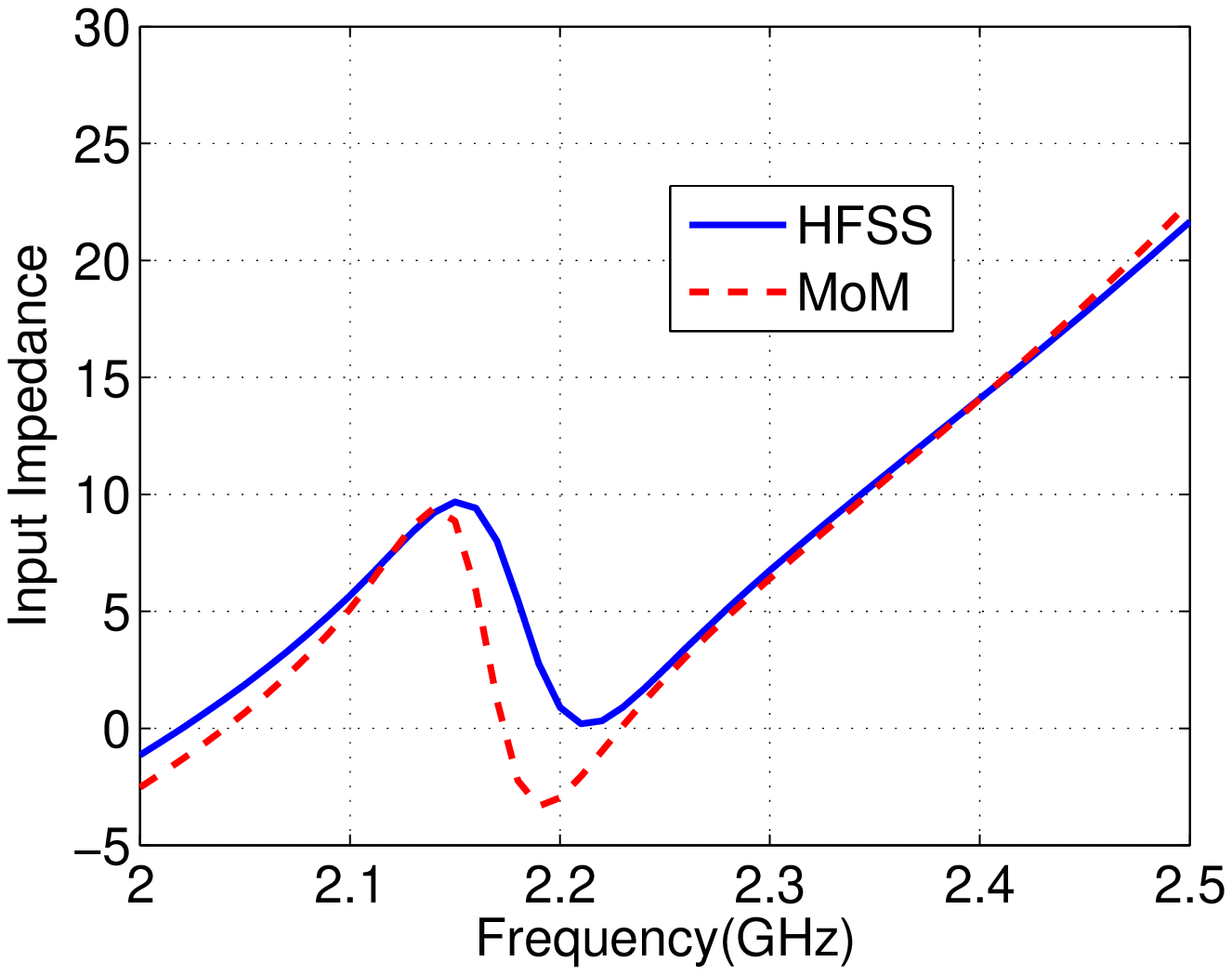}}}
  \caption{(a) Scattering term for the first three modes. (b) The input resistance and reactance. The antenna composed of a dipole of length $2l=9mm$, located at the center of a spherical core-shell resonator. Where $a_1=6.5~mm$,$a_2=7.5~mm$, $\epsilon_{r1}=60$ and $\mu_{r2}=90$.}
    \label{fig:shell2}
\end{figure}

\begin{figure}[htbp]
  \centering,
           \subfigure[]{\scalebox{0.45}{\epsfig{file=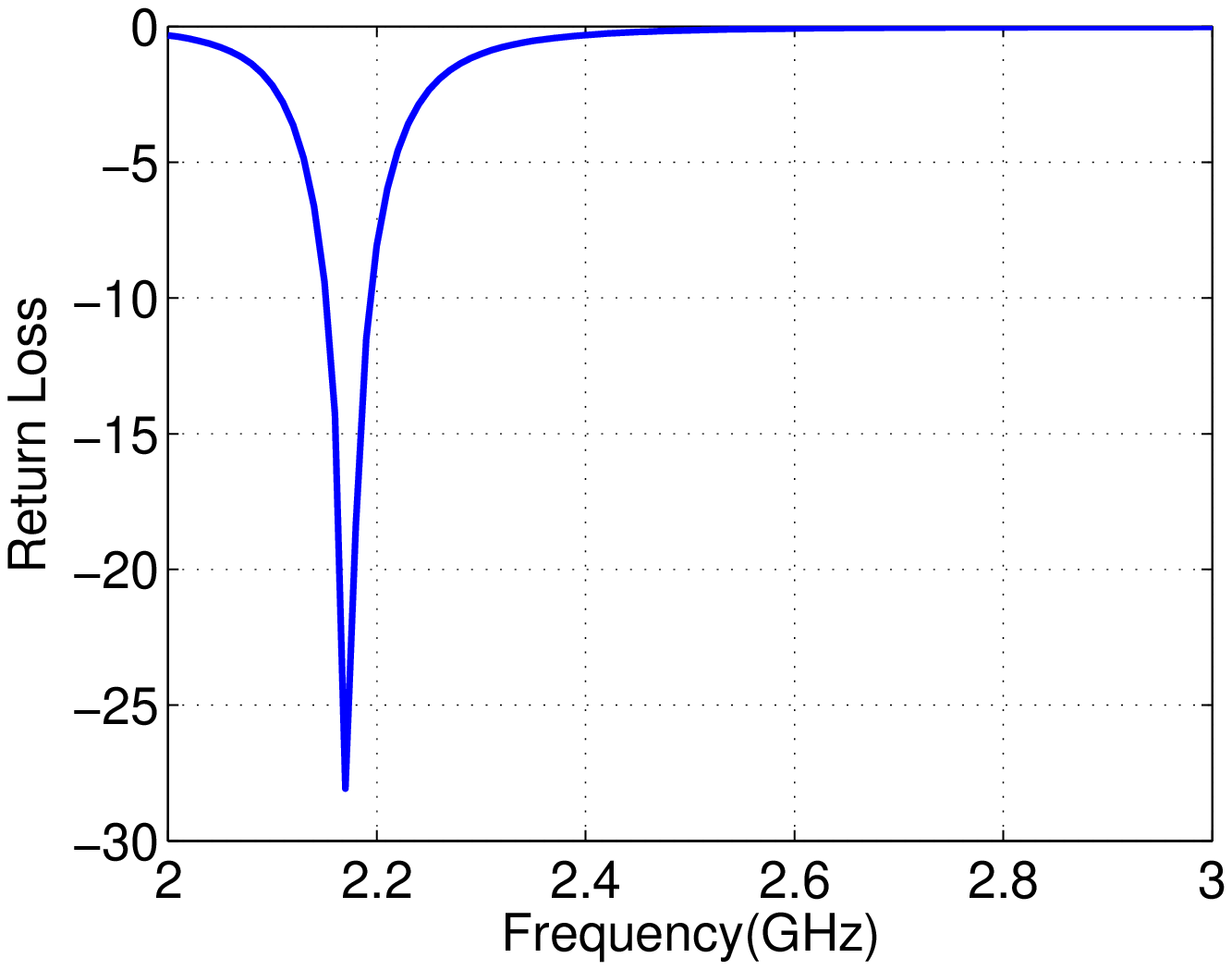}}}
            \hspace{0.2in}
           \subfigure[]{\scalebox{0.4}{\epsfig{file=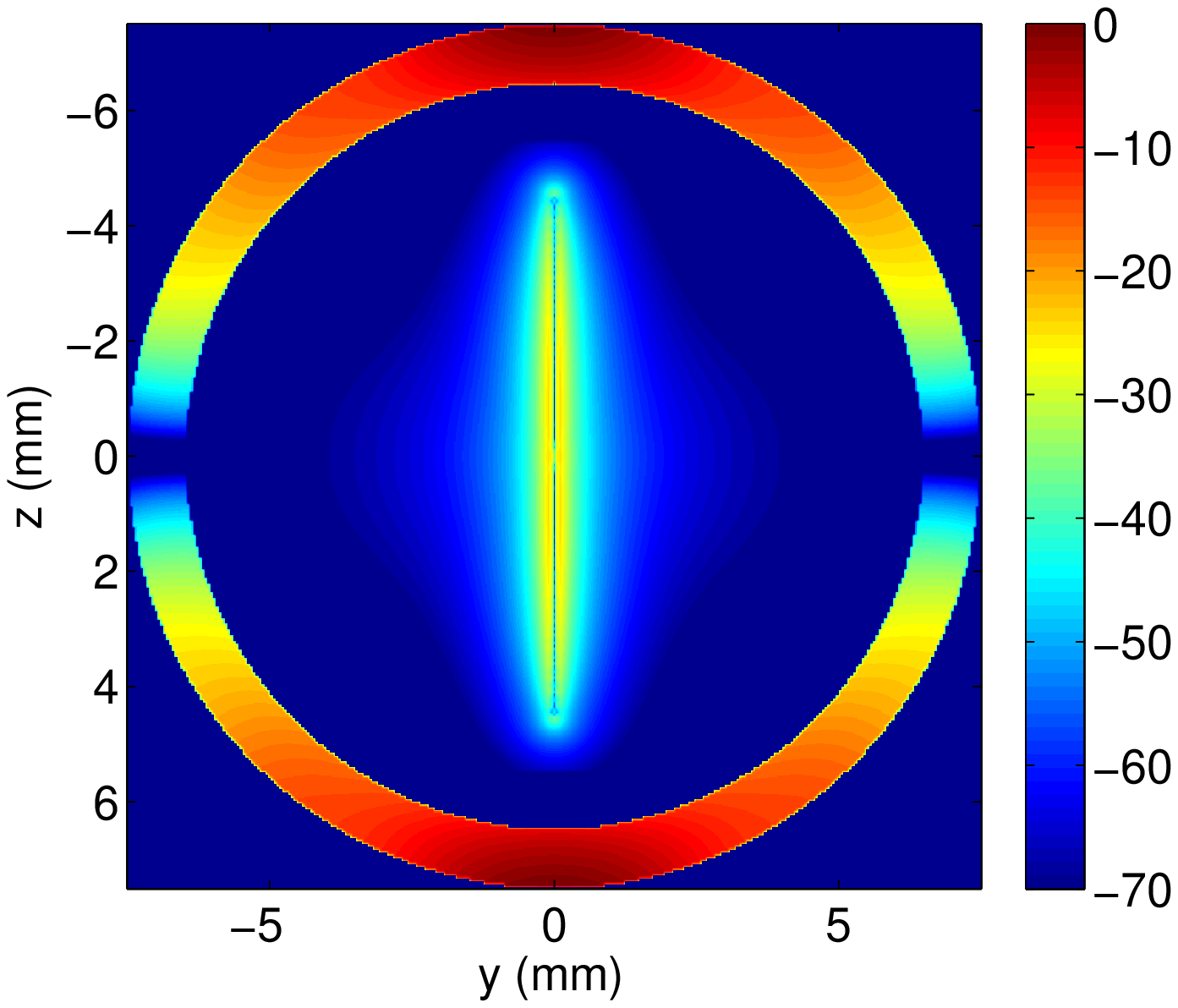}}}
  \caption{The return loss and the magnitude of electric field inside the resonator of the antenna in Fig.~\ref{fig:shell2}. The return loss presents a relatively large-bandwidths which is consistent with the $Q=31.07=1.08Q_{chu}$.}
    \label{fig:return_shell2}
\end{figure}

\quad According to Figs.~\ref{fig:scat_term}, if the dielectric material of the cores increases the bandwidths of the scattering term increases.
However, from Fig.~\ref{fig:scat_term}(b) it can be seen that the scattering terms does not varies when the dielectric material of the core is sufficiently large (in this case around $\epsilon_{r1}=60$).
Considering the scattering terms in Figs.~\ref{fig:scat_term} we can optimize a design to achieve a quality factor closer to $Q_{Chu}$ and a resonant around $2~GHz$. Fig.~\ref{fig:shell2} demonstrates the scattering terms and the input impedance for a dipole of length $2l=9~mm$ embedded inside a core-shell. The inner and outer radii of the core are $a_1=6.5~mm$ and $a_2=7.5~mm$. The core is filled with a dielectric with permittivity of $\epsilon_{r1}=60$, while the shell is a magnetic material of $\mu_{r2}=90$. The scattering term shows a significant increase in the bandwidths of the resonance. Hence, we expect a  higher bandwidths and lower Q for the antenna.
To check the accuracy of our method, we compare the input impedance vs frequency with the use of MoM and commercially available software HFSS. A very good comparison is achieved validating our method. The return loss is shown in Fig.~\ref{fig:return_shell2} exhibiting a resonant around $f=2.16~GHz$ with a large bandwidths performance. The quality factor obtained from \eqref{eq:q1}, is $Q=31.07=1.08Q_{Chu}$. The achieved results exhibit a bandwidth performance so close to Chu limit.

\begin{figure}[htbp]
  \centering,
           \subfigure[]{\scalebox{0.9}{\epsfig{file=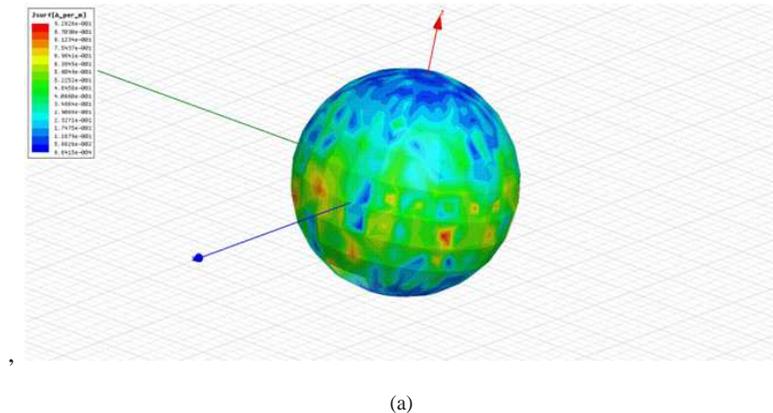}}}
  \caption{Magnitude of the surface current for the antenna in Fig.~\ref{fig:shell2}.}
    \label{fig:current}
\end{figure}

\quad To have a clear understanding of the performance the electric field inside the resonator is plotted in Fig.~\ref{fig:return_shell2}(b). The near field reveals that the magnitude of electric field inside the core is negligible (around $70~dB$ less) compared to the field inside the shell and the dipole. This means we can ignore the stored energy inside the core. Moreover, since the shell is thin, the stored energy inside the shell is small and a quality factor close to Chu can be obtained. It is worth highlighting that since the materials inside the resonator are lossless, in calculation of Q, the loss energy is zero.
Further, the magnetic shell can be envisioned as a magnetic current, which creates a kind of loop performance for the surface current and cancels the the electric current of the dipole. The magnitude of the surface current is plotted in Fig.~\ref{fig:current}. As can be seen the magnitude of the current is maximum around $\theta=90^o$ which shows the sort of loop performance (magnetic current) for the current.
The results verifies that if we have a design where the absorbed energy inside the sphere surrounding the antenna, a wide-band performance associated with a Q close to Chu limit can be achieved. The obtained observations may provide road maps for designing ultra wide-band electrically small antennas.

\quad From practical points of view, obviously, one should also consider the availability of metamaterials. For instance, the naturally available magnetic materials have always a narrow-band resonant behavior. Also as mentioned it is hard to obtain a  high $\mu$ bulk material in GHz but for thin shell it is possible to achieve a high $\mu$ material.s

\section{Conclusion}
In this paper, we investigate the behaviors of small resonators composed embedded in layered spherical resonators
by obtaining the Green's functions for the electric field. We present a novel approach for achieving the desired miniaturization while maintains the low Q performance, by embedding the dipole antenna inside a core-shell with magnetic shells.
The quality factor and input impedance of the antenna is investigated through the method of moment with Galerkin's procedure.
We illustrate that by proper choosing of the materials, quality factor as low as 1.08 times the Chu limit can be achieved.
\bibliography{LaTeX1}

\begin{thebibliography}{10}

\bibitem{chu}
L.~J. Chu.
\newblock Physical limitation on omni-directional antennas.
\newblock {\em J. of Applied Physics}, 19:1163--1175, 1948.

\bibitem{ref1}
R.~Azadegan and K.~Sarabandi.
\newblock A novel approach for miniaturization of slot antennas.
\newblock {\em IEEE Transactions on Antennas and Propagation}, 51(3):421–--429,
  2003.

\bibitem{ref2}
N.~Behdad and K.~Sarabandi.
\newblock A compact antenna for ultrawide-band applications.
\newblock {\em IEEE Transactions on Antennas and Propagation},
  53(7):2185–--2192, 2005.

\bibitem{ref4}
H.~Mosallaei and K.~Sarabandi.
\newblock Design and modeling of patch antenna printed on magneto-dielectric
  embedded-circuit metasubstrate.
\newblock {\em IEEE Transactions on Antennas and Propagation}, 55(1):45–--52,
  2007.

\bibitem{hossein}
H.~Mosallaei and Y.~Rahmat-Samii.
\newblock Broadband characterization of complex periodic ebg structures: An
  fdtd/prony technique based on the split-field approach.
\newblock {\em Electromag. J.}, 23(2):135--151, 2003.

\bibitem{stuart}
H.~R. Stuart and A.~Pidwerbetsky.
\newblock Electrically small antenna elements using negative permittivity
  resonators.
\newblock {\em IEEE Trans. Antennas Propagat.}, 54:1644--1653, 2006.

\bibitem{conf6}
H.~Mosallaei and S.~Ghadarghadr.
\newblock Negative material parameters for small antennas design.
\newblock Ottawa, Canada, 2007. URSI National Radio Science Meeting Boulder.

\bibitem{conf7}
S.~Ghadarghadr and H.~Mosallaei.
\newblock Electrically small antennas embedded in metamaterials: Closed-form
  analysis and physical insight.
\newblock Honolulu, HI, 2007. IEEE Antennas and Propagation International
  Symposium.

\bibitem{conf8}
S.~Ghadarghadr and H.~Mosallaei.
\newblock Characterization of metamaterial-based electrically small antennas.
\newblock Honolulu, HI, 2007. IEEE Antennas and Propagation International
  Symposium.

\bibitem{j3}
S.~Ghadarghadr, A.~Ahmadi, and H.~Mosallaei.
\newblock Negative permeability-based electrically small antennas.
\newblock {\em IEEE Transactions on Antennas and Wireless Propagation Letters},
  7:13--17, 2008.

\bibitem{zil}
R.~W. Ziolkowski and A.~D. Kipple.
\newblock Application of aouble negative materials to increase the power
  radiated by electrically small antennas.
\newblock {\em IEEE Trans. Antennas Propagat.}, 51:2626--2640, 2003.

\bibitem{harrington}
R.~F. Harrington.
\newblock {\em Time Harmonic Electromagnetic Fields}.
\newblock McGraw-Hill, New York, 1961.

\bibitem{chew}
W.~C. Chew.
\newblock {\em Waves and Fields in Inhomogeneous Media}.
\newblock IEEE Press, New York, 1995.

\bibitem{lin}
K.~W. Leung, K.~M. Luk, K.~Y.~A. Lai, and D.~Lin.
\newblock Theory and experiment of a coaxial probe fed hemispherical dielectric
  resonator antenna.
\newblock {\em IEEE Trans. Antennas Propagat.}, 41:1390--1398, 1993.

\bibitem{yaghiq}
A.~D. Yaghjian and S.~R. Best.
\newblock Impedance, bandwidth and q of antennas.
\newblock {\em IEEE Transactions on Antennas and Propagation}, 53(4):1298--
  1324, 2005.

\end{thebibliography}
\bibliographystyle{unsrt}
\end{document}